\documentclass[preprint,prd,tightenlines,superscriptaddress]{revtex4-1}

\usepackage{graphicx} % Include figure files
\usepackage{dcolumn}  % Align table columns on decimal point
\usepackage{colordvi}
\usepackage{color}
\usepackage{epstopdf}
\usepackage{pstricks}
\usepackage{amssymb}
\usepackage{url}
\graphicspath{{ps}}
\usepackage{hyperref}
\usepackage{tabularx}
\usepackage{multirow}
\usepackage{units}
\usepackage{siunitx}
\usepackage{hyphenat}
\usepackage{subfigure}
\usepackage[italic]{hepnames}
\usepackage{upgreek}
\usepackage{xspace}
\usepackage{color, colortbl}

\renewcommand{\arraystretch}{1.1}

\newcommand{\ee}{\text{e}}

\newcommand{\CP}{\ensuremath{C\hspace{-0.13em}P}\xspace}

\renewcommand{\Ph}{\HepParticle{h}{}{}\xspace}

\newcommand{\phione}{\ensuremath{\upphi_1}\xspace}
\newcommand{\phitwo}{\ensuremath{\upphi_2}\xspace}

\renewcommand{\PUpsilonFourS}{\HepParticleResonance{\PgU}{4\mathrm{S}}{}{}\xspace}
\newcommand{\Pa}{\HepParticle{a}{}{}\xspace}

\renewcommand{\PBzero}{\ensuremath{\HepParticle{\PB}{}{}^0}\xspace}
\renewcommand{\APBzero}{\ensuremath{\HepParticle{\APB}{}{}^0}\xspace}
\renewcommand{\APDzero}{\ensuremath{\HepParticle{\APD}{}{}^0}\xspace}
\renewcommand{\Pgpz}{\ensuremath{\HepParticle{\Pgp}{}{}^0}\xspace}
\renewcommand{\PDzero}{\ensuremath{\HepParticle{\PD}{}{}^0}\xspace}
\renewcommand{\PKzS}{\ensuremath{\HepParticle{\PK}{}{}^0_{\rm S}}\xspace}

\renewcommand{\PUpsilonFourS}{\HepParticleResonance{\PgU}{4\mathrm{S}}{}{}\xspace}

\begin{document}

%place for definitions and newcommands
\def\belletwo {\it {Belle II}}

%% Avoid "orphans" and "widows"
\clubpenalty = 10000  % no orphans
\widowpenalty = 10000 % no widows

\vspace*{-3\baselineskip}
\resizebox{!}{3cm}{\includegraphics{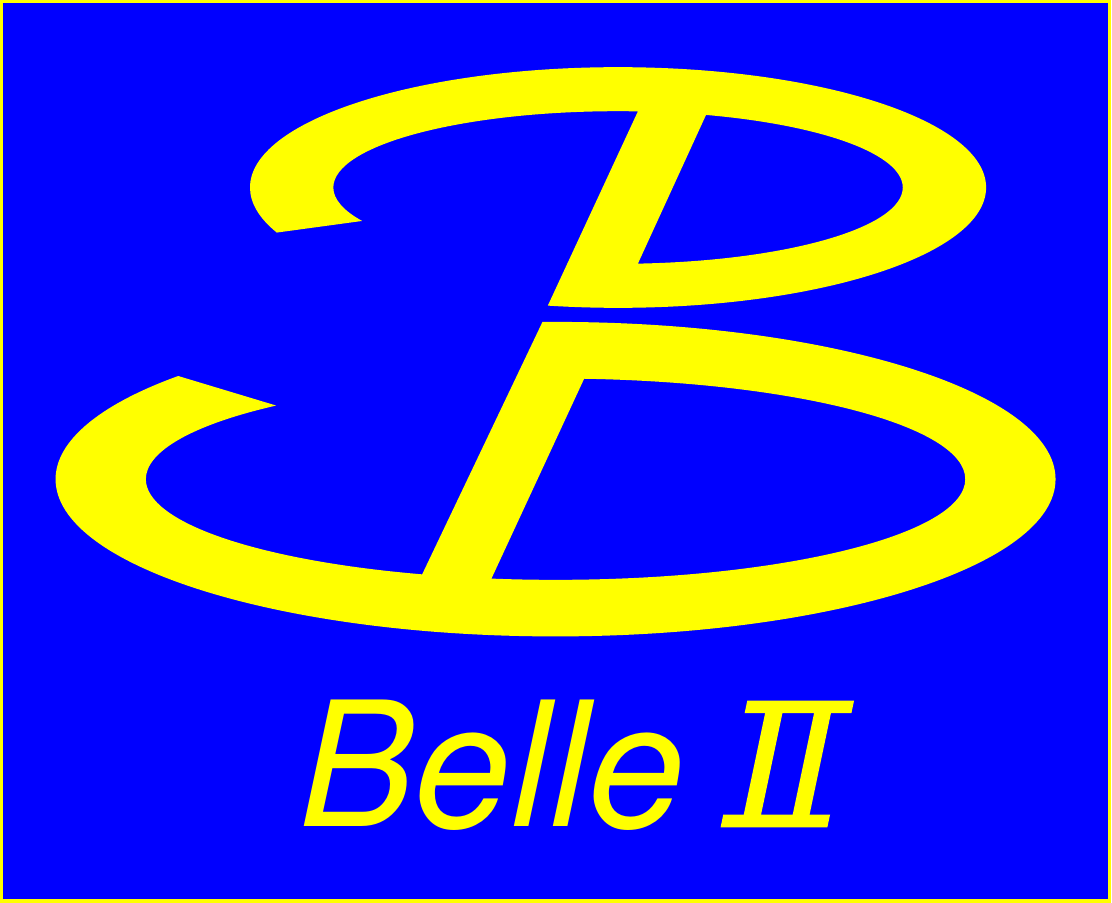}}

\vspace*{-5\baselineskip}
\begin{flushright}
% BELLE2-CONF-DRAFT-2020-018
BELLE2-CONF-PH-2020-004
\\
%Version 3.0 \\
\today
\end{flushright}

\quad\\[0.5cm]

\title {First flavor tagging calibration using 2019 Belle II data}

%%% Paper:    (2020  conference papers)
%%% Journal:  (2020 conferences)
%%% ====================================================================
%%% Use \input{authors-conf2020} to insert this material into your latex file.
\newcommand{\instSinica}{Academia Sinica, Taipei 11529, Taiwan}
\newcommand{\instCPPM}{Aix Marseille Universit\'{e}, CNRS/IN2P3, CPPM, 13288 Marseille, France}
\newcommand{\instBeihang}{Beihang University, Beijing 100191, China}
\newcommand{\instBUAP}{Benemerita Universidad Autonoma de Puebla, Puebla 72570, Mexico}
\newcommand{\instBNL}{Brookhaven National Laboratory, Upton, New York 11973, U.S.A.}
\newcommand{\instBINP}{Budker Institute of Nuclear Physics SB RAS, Novosibirsk 630090, Russian Federation}
\newcommand{\instCMU}{Carnegie Mellon University, Pittsburgh, Pennsylvania 15213, U.S.A.}
\newcommand{\instCinvestavIPN}{Centro de Investigacion y de Estudios Avanzados del Instituto Politecnico Nacional, Mexico City 07360, Mexico}
\newcommand{\instPrague}{Faculty of Mathematics and Physics, Charles University, 121 16 Prague, Czech Republic}
\newcommand{\instChiangMai}{Chiang Mai University, Chiang Mai 50202, Thailand}
\newcommand{\instChiba}{Chiba University, Chiba 263-8522, Japan}
\newcommand{\instChonnam}{Chonnam National University, Gwangju 61186, South Korea}
\newcommand{\instConacyt}{Consejo Nacional de Ciencia y Tecnolog\'{\i}a, Mexico City 03940, Mexico}
\newcommand{\instDESY}{Deutsches Elektronen--Synchrotron, 22607 Hamburg, Germany}
\newcommand{\instDuke}{Duke University, Durham, North Carolina 27708, U.S.A.}
\newcommand{\instITAR}{Institute of Theoretical and Applied Research (ITAR), Duy Tan University, Hanoi 100000, Vietnam}
\newcommand{\instENEA}{ENEA Casaccia, I-00123 Roma, Italy}
\newcommand{\instEri}{Earthquake Research Institute, University of Tokyo, Tokyo 113-0032, Japan}
\newcommand{\instJuelich}{Forschungszentrum J\"{u}lich, 52425 J\"{u}lich, Germany}
\newcommand{\instFuJen}{Department of Physics, Fu Jen Catholic University, Taipei 24205, Taiwan}
\newcommand{\instFudan}{Key Laboratory of Nuclear Physics and Ion-beam Application (MOE) and Institute of Modern Physics, Fudan University, Shanghai 200443, China}
\newcommand{\instGoettingen}{II. Physikalisches Institut, Georg-August-Universit\"{a}t G\"{o}ttingen, 37073 G\"{o}ttingen, Germany}
\newcommand{\instGifu}{Gifu University, Gifu 501-1193, Japan}
\newcommand{\instSOKENDAI}{The Graduate University for Advanced Studies (SOKENDAI), Hayama 240-0193, Japan}
\newcommand{\instGyeongsang}{Gyeongsang National University, Jinju 52828, South Korea}
\newcommand{\instHanyang}{Department of Physics and Institute of Natural Sciences, Hanyang University, Seoul 04763, South Korea}
\newcommand{\instKEK}{High Energy Accelerator Research Organization (KEK), Tsukuba 305-0801, Japan}
\newcommand{\instJPARC}{J-PARC Branch, KEK Theory Center, High Energy Accelerator Research Organization (KEK), Tsukuba 305-0801, Japan}
\newcommand{\instHSE}{Higher School of Economics (HSE), Moscow 101000, Russian Federation}
\newcommand{\instIISER}{Indian Institute of Science Education and Research Mohali, SAS Nagar, 140306, India}
\newcommand{\instIITBhubaneswar}{Indian Institute of Technology Bhubaneswar, Satya Nagar 751007, India}
\newcommand{\instIITGuwahati}{Indian Institute of Technology Guwahati, Assam 781039, India}
\newcommand{\instIITHyderabad}{Indian Institute of Technology Hyderabad, Telangana 502285, India}
\newcommand{\instIITMadras}{Indian Institute of Technology Madras, Chennai 600036, India}
\newcommand{\instIndiana}{Indiana University, Bloomington, Indiana 47408, U.S.A.}
\newcommand{\instIHEPRussia}{Institute for High Energy Physics, Protvino 142281, Russian Federation}
\newcommand{\instHEPHYVienna}{Institute of High Energy Physics, Vienna 1050, Austria}
\newcommand{\instIHEPChina}{Institute of High Energy Physics, Chinese Academy of Sciences, Beijing 100049, China}
\newcommand{\instChennai}{Institute of Mathematical Sciences, Chennai 600113, India}
\newcommand{\instIPP}{Institute of Particle Physics (Canada), Victoria, British Columbia V8W 2Y2, Canada}
\newcommand{\instIOP}{Institute of Physics, Vietnam Academy of Science and Technology (VAST), Hanoi, Vietnam}
\newcommand{\instIFIC}{Instituto de Fisica Corpuscular, Paterna 46980, Spain}
\newcommand{\instFrascati}{INFN Laboratori Nazionali di Frascati, I-00044 Frascati, Italy}
\newcommand{\instNapoliINFN}{INFN Sezione di Napoli, I-80126 Napoli, Italy}
\newcommand{\instPadovaINFN}{INFN Sezione di Padova, I-35131 Padova, Italy}
\newcommand{\instPerugiaINFN}{INFN Sezione di Perugia, I-06123 Perugia, Italy}
\newcommand{\instPisaINFN}{INFN Sezione di Pisa, I-56127 Pisa, Italy}
\newcommand{\instRomaINFN}{INFN Sezione di Roma, I-00185 Roma, Italy}
\newcommand{\instRomaTreINFN}{INFN Sezione di Roma Tre, I-00146 Roma, Italy}
\newcommand{\instTorinoINFN}{INFN Sezione di Torino, I-10125 Torino, Italy}
\newcommand{\instTriesteINFN}{INFN Sezione di Trieste, I-34127 Trieste, Italy}
\newcommand{\instJAEA}{Advanced Science Research Center, Japan Atomic Energy Agency, Naka 319-1195, Japan}
\newcommand{\instMainz}{Johannes Gutenberg-Universit\"{a}t Mainz, Institut f\"{u}r Kernphysik, D-55099 Mainz, Germany}
\newcommand{\instGiessen}{Justus-Liebig-Universit\"{a}t Gie\ss{}en, 35392 Gie\ss{}en, Germany}
\newcommand{\instKarlsruhe}{Institut f\"{u}r Experimentelle Teilchenphysik, Karlsruher Institut f\"{u}r Technologie, 76131 Karlsruhe, Germany}
\newcommand{\instKennesaw}{Kennesaw State University, Kennesaw, Georgia 30144, U.S.A.}
\newcommand{\instKitasato}{Kitasato University, Sagamihara 252-0373, Japan}
\newcommand{\instKISTI}{Korea Institute of Science and Technology Information, Daejeon 34141, South Korea}
\newcommand{\instKorea}{Korea University, Seoul 02841, South Korea}
\newcommand{\instKSU}{Kyoto Sangyo University, Kyoto 603-8555, Japan}
\newcommand{\instKyotoU}{Kyoto University, Kyoto 606-8501, Japan}
\newcommand{\instKyungpook}{Kyungpook National University, Daegu 41566, South Korea}
\newcommand{\instLPI}{P.N. Lebedev Physical Institute of the Russian Academy of Sciences, Moscow 119991, Russian Federation}
\newcommand{\instLNNU}{Liaoning Normal University, Dalian 116029, China}
\newcommand{\instLMU}{Ludwig Maximilians University, 80539 Munich, Germany}
\newcommand{\instISUni}{Iowa State University,  Ames, Iowa 50011, U.S.A.}
\newcommand{\instLuther}{Luther College, Decorah, Iowa 52101, U.S.A.}
\newcommand{\instMNITJaipur}{Malaviya National Institute of Technology Jaipur, Jaipur 302017, India}
\newcommand{\instMPP}{Max-Planck-Institut f\"{u}r Physik, 80805 M\"{u}nchen, Germany}
\newcommand{\instMPGHLL}{Semiconductor Laboratory of the Max Planck Society, 81739 M\"{u}nchen, Germany}
\newcommand{\instMcGill}{McGill University, Montr\'{e}al, Qu\'{e}bec, H3A 2T8, Canada}
\newcommand{\instMETU}{Middle East Technical University, 06531 Ankara, Turkey}
\newcommand{\instMEPhI}{Moscow Physical Engineering Institute, Moscow 115409, Russian Federation}
\newcommand{\instNagoya}{Graduate School of Science, Nagoya University, Nagoya 464-8602, Japan}
\newcommand{\instNagoyaKMI}{Kobayashi-Maskawa Institute, Nagoya University, Nagoya 464-8602, Japan}
\newcommand{\instNagoyaIAR}{Institute for Advanced Research, Nagoya University, Nagoya 464-8602, Japan}
\newcommand{\instNaraWu}{Nara Women's University, Nara 630-8506, Japan}
\newcommand{\instUNAM}{National Autonomous University of Mexico, Mexico City, Mexico}
\newcommand{\instNTUTaiwan}{Department of Physics, National Taiwan University, Taipei 10617, Taiwan}
\newcommand{\instNUUTaiwan}{National United University, Miao Li 36003, Taiwan}
\newcommand{\instKrakow}{H. Niewodniczanski Institute of Nuclear Physics, Krakow 31-342, Poland}
\newcommand{\instNiigata}{Niigata University, Niigata 950-2181, Japan}
\newcommand{\instNSU}{Novosibirsk State University, Novosibirsk 630090, Russian Federation}
\newcommand{\instOkinawa}{Okinawa Institute of Science and Technology, Okinawa 904-0495, Japan}
\newcommand{\instOsakaCity}{Osaka City University, Osaka 558-8585, Japan}
\newcommand{\instRCNP}{Research Center for Nuclear Physics, Osaka University, Osaka 567-0047, Japan}
\newcommand{\instPNNL}{Pacific Northwest National Laboratory, Richland, Washington 99352, U.S.A.}
\newcommand{\instPanjab}{Panjab University, Chandigarh 160014, India}
\newcommand{\instPeking}{Peking University, Beijing 100871, China}
\newcommand{\instPanjabPAU}{Punjab Agricultural University, Ludhiana 141004, India}
\newcommand{\instRIKENMSL}{Meson Science Laboratory, Cluster for Pioneering Research, RIKEN, Saitama 351-0198, Japan}
\newcommand{\instRIKEN}{Theoretical Research Division, Nishina Center, RIKEN, Saitama 351-0198, Japan}
\newcommand{\instXavier}{St. Francis Xavier University, Antigonish, Nova Scotia, B2G 2W5, Canada}
\newcommand{\instSeoul}{Seoul National University, Seoul 08826, South Korea}
\newcommand{\instShandong}{Shandong University, Jinan 250100, China}
\newcommand{\instSPU}{Showa Pharmaceutical University, Tokyo 194-8543, Japan}
\newcommand{\instSoochow}{Soochow University, Suzhou 215006, China}
\newcommand{\instSoongsil}{Soongsil University, Seoul 06978, South Korea}
\newcommand{\instLjubljanaJSI}{J. Stefan Institute, 1000 Ljubljana, Slovenia}
\newcommand{\instKyiv}{Taras Shevchenko National Univ. of Kiev, Kiev, Ukraine}
\newcommand{\instTata}{Tata Institute of Fundamental Research, Mumbai 400005, India}
\newcommand{\instTUM}{Department of Physics, Technische Universit\"{a}t M\"{u}nchen, 85748 Garching, Germany}
\newcommand{\instECUTUM}{Excellence Cluster Universe, Technische Universit\"{a}t M\"{u}nchen, 85748 Garching, Germany}
\newcommand{\instTelAviv}{Tel Aviv University, School of Physics and Astronomy, Tel Aviv, 69978, Israel}
\newcommand{\instToho}{Toho University, Funabashi 274-8510, Japan}
\newcommand{\instTohoku}{Department of Physics, Tohoku University, Sendai 980-8578, Japan}
\newcommand{\instTitech}{Tokyo Institute of Technology, Tokyo 152-8550, Japan}
\newcommand{\instTokyoMetropolitan}{Tokyo Metropolitan University, Tokyo 192-0397, Japan}
\newcommand{\instUAS}{Universidad Autonoma de Sinaloa, Sinaloa 80000, Mexico}
\newcommand{\instNapoliUNIV}{Dipartimento di Scienze Fisiche, Universit\`{a} di Napoli Federico II, I-80126 Napoli, Italy}
\newcommand{\instNapoliUNIVA}{Dipartimento di Agraria, Universit\`{a} di Napoli Federico II, I-80055 Portici (NA), Italy}
\newcommand{\instPadovaUNIV}{Dipartimento di Fisica e Astronomia, Universit\`{a} di Padova, I-35131 Padova, Italy}
\newcommand{\instPerugiaUNIV}{Dipartimento di Fisica, Universit\`{a} di Perugia, I-06123 Perugia, Italy}
\newcommand{\instPisaUNIV}{Dipartimento di Fisica, Universit\`{a} di Pisa, I-56127 Pisa, Italy}
\newcommand{\instRomaUNIV}{Universit\`{a} di Roma ``La Sapienza,'' I-00185 Roma, Italy}
\newcommand{\instRomaTreUNIV}{Dipartimento di Matematica e Fisica, Universit\`{a} di Roma Tre, I-00146 Roma, Italy}
\newcommand{\instTorinoUNIV}{Dipartimento di Fisica, Universit\`{a} di Torino, I-10125 Torino, Italy}
\newcommand{\instTriesteUNIV}{Dipartimento di Fisica, Universit\`{a} di Trieste, I-34127 Trieste, Italy}
\newcommand{\instMontreal}{Universit\'{e} de Montr\'{e}al, Physique des Particules, Montr\'{e}al, Qu\'{e}bec, H3C 3J7, Canada}
\newcommand{\instIJCLab}{Universit\'{e} Paris-Saclay, CNRS/IN2P3, IJCLab, 91405 Orsay, France}
\newcommand{\instIPHC}{Universit\'{e} de Strasbourg, CNRS, IPHC, UMR 7178, 67037 Strasbourg, France}
\newcommand{\instAdelaide}{Department of Physics, University of Adelaide, Adelaide, South Australia 5005, Australia}
\newcommand{\instBonn}{University of Bonn, 53115 Bonn, Germany}
\newcommand{\instUBC}{University of British Columbia, Vancouver, British Columbia, V6T 1Z1, Canada}
\newcommand{\instCincinnati}{University of Cincinnati, Cincinnati, Ohio 45221, U.S.A.}
\newcommand{\instFlorida}{University of Florida, Gainesville, Florida 32611, U.S.A.}
\newcommand{\instHamburg}{University of Hamburg, 20148 Hamburg, Germany}
\newcommand{\instHawaii}{University of Hawaii, Honolulu, Hawaii 96822, U.S.A.}
\newcommand{\instHeidelberg}{University of Heidelberg, 68131 Mannheim, Germany}
\newcommand{\instLjubljanaUniLJ}{Faculty of Mathematics and Physics, University of Ljubljana, 1000 Ljubljana, Slovenia}
\newcommand{\instLouisville}{University of Louisville, Louisville, Kentucky 40292, U.S.A.}
\newcommand{\instMalaya}{National Centre for Particle Physics, University Malaya, 50603 Kuala Lumpur, Malaysia}
\newcommand{\instLjubljanaUM}{University of Maribor, 2000 Maribor, Slovenia}
\newcommand{\instMelbourne}{School of Physics, University of Melbourne, Victoria 3010, Australia}
\newcommand{\instMississippi}{University of Mississippi, University, Mississippi 38677, U.S.A.}
\newcommand{\instUOM}{University of Miyazaki, Miyazaki 889-2192, Japan}
\newcommand{\instNovaGorica}{University of Nova Gorica, 5000 Nova Gorica, Slovenia}
\newcommand{\instPittsburgh}{University of Pittsburgh, Pittsburgh, Pennsylvania 15260, U.S.A.}
\newcommand{\instUSTC}{University of Science and Technology of China, Hefei 230026, China}
\newcommand{\instSAlabama}{University of South Alabama, Mobile, Alabama 36688, U.S.A.}
\newcommand{\instSCarolina}{University of South Carolina, Columbia, South Carolina 29208, U.S.A.}
\newcommand{\instSydney}{School of Physics, University of Sydney, New South Wales 2006, Australia}
\newcommand{\instTabuk}{Department of Physics, Faculty of Science, University of Tabuk, Tabuk 71451, Saudi Arabia}
\newcommand{\instUTokyo}{Department of Physics, University of Tokyo, Tokyo 113-0033, Japan}
\newcommand{\instIPMU}{Kavli Institute for the Physics and Mathematics of the Universe (WPI), University of Tokyo, Kashiwa 277-8583, Japan}
\newcommand{\instVictoria}{University of Victoria, Victoria, British Columbia, V8W 3P6, Canada}
\newcommand{\instVPI}{Virginia Polytechnic Institute and State University, Blacksburg, Virginia 24061, U.S.A.}
\newcommand{\instWayneState}{Wayne State University, Detroit, Michigan 48202, U.S.A.}
\newcommand{\instYamagata}{Yamagata University, Yamagata 990-8560, Japan}
\newcommand{\instYerevan}{Alikhanyan National Science Laboratory, Yerevan 0036, Armenia}
\newcommand{\instYonsei}{Yonsei University, Seoul 03722, South Korea}
%%%\affiliation{\instSinica}
\affiliation{\instCPPM}
\affiliation{\instBeihang}
%%%\affiliation{\instBUAP}
\affiliation{\instBNL}
\affiliation{\instBINP}
\affiliation{\instCMU}
\affiliation{\instCinvestavIPN}
\affiliation{\instPrague}
\affiliation{\instChiangMai}
\affiliation{\instChiba}
\affiliation{\instChonnam}
\affiliation{\instConacyt}
\affiliation{\instDESY}
\affiliation{\instDuke}
\affiliation{\instITAR}
%%%\affiliation{\instENEA}
\affiliation{\instEri}
\affiliation{\instJuelich}
\affiliation{\instFuJen}
\affiliation{\instFudan}
\affiliation{\instGoettingen}
\affiliation{\instGifu}
\affiliation{\instSOKENDAI}
\affiliation{\instGyeongsang}
\affiliation{\instHanyang}
\affiliation{\instKEK}
\affiliation{\instJPARC}
\affiliation{\instHSE}
\affiliation{\instIISER}
\affiliation{\instIITBhubaneswar}
\affiliation{\instIITGuwahati}
\affiliation{\instIITHyderabad}
\affiliation{\instIITMadras}
\affiliation{\instIndiana}
\affiliation{\instIHEPRussia}
\affiliation{\instHEPHYVienna}
\affiliation{\instIHEPChina}
%%%\affiliation{\instChennai}
\affiliation{\instIPP}
\affiliation{\instIOP}
\affiliation{\instIFIC}
\affiliation{\instFrascati}
\affiliation{\instNapoliINFN}
\affiliation{\instPadovaINFN}
\affiliation{\instPerugiaINFN}
\affiliation{\instPisaINFN}
\affiliation{\instRomaINFN}
\affiliation{\instRomaTreINFN}
\affiliation{\instTorinoINFN}
\affiliation{\instTriesteINFN}
\affiliation{\instISUni}
\affiliation{\instJAEA}
\affiliation{\instMainz}
\affiliation{\instGiessen}
\affiliation{\instKarlsruhe}
%%%\affiliation{\instKennesaw}
\affiliation{\instKitasato}
\affiliation{\instKISTI}
\affiliation{\instKorea}
\affiliation{\instKSU}
%%%\affiliation{\instKyotoU}
\affiliation{\instKyungpook}
\affiliation{\instLPI}
\affiliation{\instLNNU}
\affiliation{\instLMU}
\affiliation{\instLuther}
\affiliation{\instMNITJaipur}
\affiliation{\instMPP}
\affiliation{\instMPGHLL}
\affiliation{\instMcGill}
%%%\affiliation{\instMETU}
\affiliation{\instMEPhI}
\affiliation{\instNagoya}
\affiliation{\instNagoyaKMI}
\affiliation{\instNagoyaIAR}
\affiliation{\instNaraWu}
%%%\affiliation{\instUNAM}
\affiliation{\instNTUTaiwan}
\affiliation{\instNUUTaiwan}
\affiliation{\instKrakow}
\affiliation{\instNiigata}
\affiliation{\instNSU}
\affiliation{\instOkinawa}
\affiliation{\instOsakaCity}
\affiliation{\instRCNP}
\affiliation{\instPNNL}
\affiliation{\instPanjab}
\affiliation{\instPeking}
\affiliation{\instPanjabPAU}
\affiliation{\instRIKENMSL}
%%%\affiliation{\instRIKEN}
%%%\affiliation{\instXavier}
\affiliation{\instSeoul}
%%%\affiliation{\instShandong}
\affiliation{\instSPU}
\affiliation{\instSoochow}
\affiliation{\instSoongsil}
\affiliation{\instLjubljanaJSI}
\affiliation{\instKyiv}
\affiliation{\instTata}
\affiliation{\instTUM}
%%%\affiliation{\instECUTUM}
\affiliation{\instTelAviv}
\affiliation{\instToho}
\affiliation{\instTohoku}
\affiliation{\instTitech}
\affiliation{\instTokyoMetropolitan}
\affiliation{\instUAS}
\affiliation{\instNapoliUNIV}
\affiliation{\instPadovaUNIV}
\affiliation{\instPerugiaUNIV}
\affiliation{\instPisaUNIV}
\affiliation{\instRomaUNIV}
\affiliation{\instRomaTreUNIV}
\affiliation{\instTorinoUNIV}
\affiliation{\instTriesteUNIV}
\affiliation{\instMontreal}
\affiliation{\instIJCLab}
\affiliation{\instIPHC}
\affiliation{\instAdelaide}
\affiliation{\instBonn}
\affiliation{\instUBC}
\affiliation{\instCincinnati}
\affiliation{\instFlorida}
%%%\affiliation{\instHamburg}
\affiliation{\instHawaii}
\affiliation{\instHeidelberg}
\affiliation{\instLjubljanaUniLJ}
\affiliation{\instLouisville}
\affiliation{\instMalaya}
\affiliation{\instLjubljanaUM}
\affiliation{\instMelbourne}
\affiliation{\instMississippi}
\affiliation{\instUOM}
%%%\affiliation{\instNovaGorica}
\affiliation{\instPittsburgh}
\affiliation{\instUSTC}
\affiliation{\instSAlabama}
\affiliation{\instSCarolina}
\affiliation{\instSydney}
%%%\affiliation{\instTabuk}
\affiliation{\instUTokyo}
\affiliation{\instIPMU}
\affiliation{\instVictoria}
\affiliation{\instVPI}
\affiliation{\instWayneState}
\affiliation{\instYamagata}
\affiliation{\instYerevan}
\affiliation{\instYonsei}
  \author{F.~Abudin{\'e}n}\affiliation{\instTriesteINFN} % 2250
  \author{I.~Adachi}\affiliation{\instKEK}\affiliation{\instSOKENDAI} % 2590
  \author{R.~Adak}\affiliation{\instFudan} % 6743
  \author{K.~Adamczyk}\affiliation{\instKrakow} % 2239
  \author{P.~Ahlburg}\affiliation{\instBonn} % 2367
  \author{J.~K.~Ahn}\affiliation{\instKorea} % 7423
  \author{H.~Aihara}\affiliation{\instUTokyo} % 2223
  \author{N.~Akopov}\affiliation{\instYerevan} % 9443
  \author{A.~Aloisio}\affiliation{\instNapoliUNIV}\affiliation{\instNapoliINFN} % 2194
  \author{F.~Ameli}\affiliation{\instRomaINFN} % 4683
% ...
% 2543
\collaboration{Belle II Collaboration}

%\noaffiliation

\begin{abstract}
We report on the first calibration of the standard Belle~II~$\PB$-flavor tagger using the full data set collected at the $\Upsilon(4{\rm S})$ resonance in 2019 with the Belle II detector at the SuperKEKB collider, corresponding to 8.7\,fb$^{-1}$ of integrated luminosity. The calibration is performed by reconstructing \mbox{various} hadronic charmed \PB-meson decays with flavor-specific final states. We use simulation to optimize our event selection criteria and to train the flavor tagging algorithm. 
%To validate the output of the algorithm and to measure the calibration constants, we reconstruct various self-tagging charmed $\PB$~decays from electron-positron collisions at the energy corresponding to the $\Upsilon(4{\rm S})$ resonance.
%  We fit the difference between half of the collision energy and the $B$ candidate energy (in the $\Upsilon(4{\rm S})$ frame) for events restricted to a signal-rich range in beam-energy-constrained mass to separate signal $\PB$ candidates from the different backgrounds. 
We determine the tagging \mbox{efficiency} and the fraction of wrongly identified tag-side \PB~candidates from a measurement of the time-integrated $B^0-\overline{B}^0$~\mbox{mixing} probability. 
The total effective efficiency is measured to be \mbox{$\varepsilon_{\rm eff} = \big(33.8 \pm 3.6(\text{stat}) \pm 1.6(\text{sys})\big)\%$},
%The total effective efficiency for neutral $\PB$~candidates is measured to be \mbox{$\varepsilon_{\rm eff} = 33.8 \pm 3.6(\text{stat}) \pm 1.6(\text{sys}) $}, and for charged $\PB$~candidates \mbox{$\varepsilon_{\rm eff} = 36.6 \pm 1.8(\text{stat}) \pm 0.7(\text{sys}) $}.  
%To cross-check the results, we measure the $\PBzero-\APBzero$~mixing probability~$\chi_{d}$. For neutral candidates, we obtain the value $\chi_d = \pm (\text{stat}) \pm (\text{sys}) $, which is in good agreement with the PDG value. For charged candidates we obtain $\chi_d = \pm (\text{stat}) \pm (\text{sys}) $, which is in good agreement with $0$. 
 which is in good agreement with the predictions from simulation and comparable with the best one obtained by the Belle experiment.
The results show a good understanding of the detector performance and offer a basis for future calibrations.

\keywords{Belle II, charmless, phase 3}
\end{abstract}

\pacs{}

\maketitle

{\renewcommand{\thefootnote}{\fnsymbol{footnote}}}
\setcounter{footnote}{0}

%\tableofcontents

%%%%%%%%%% Introduction %%%%%%%%%%%%

\section{Introduction and motivation}

Flavor tagging is the task of determining the heavy quark-flavor content of mesons. At Belle~II, determining the flavor of neutral \PB~mesons is needed for many measurements of \mbox{$\PBzero-\APBzero$}~mixing and \CP-violation, where  usually a signal $\PB$~meson is fully reconstructed~(signal side) and the flavor of the accompanying \PB-meson~(tag side) has to be determined. Thus, flavor tagging plays an essential role in precise measurements of the CKM angles $\phione/\upbeta$ and $\phitwo/\upalpha$ and in the study of flavor anomalies that could ultimately reveal possible deviations from standard model expectations. 
% Additionally, the \PB-flavor information is exploited in many analyses to provide additional discrimination power against continuum background.  

At Belle~II, flavor tagging is accomplished using multivariate approaches. The standard algorithm is a category-based flavor tagger~\cite{Abudinen:2018} that first identifies \PBzero-decay products \mbox{providing} flavor information and then combines all information to determine the \PBzero~flavor. There is another algorithm, a deep-learning flavor tagger (DNN)~\cite{Gemmler:2016}, that determines the \PBzero~flavor in a single step without pre-identifying \PBzero-decay products. The performance of this algorithm in Belle~II data is  currently being evaluated and is planned to be calibrated in the future. 

In this work, we calibrate the category-based flavor tagger by measuring the time-integrated $\PBzero-\APBzero$~mixing probability. 
%For this, we determine the flavor of the two $\PB$~mesons coming from the $\Upsilon(4{\rm S})$ decay. 
We reconstruct signal $\PB$~decays with final states that allow us to unambiguously identify the flavor of the signal side and determine the flavor of the tag side using the flavor tagger. We
reconstruct charmed signal $\PB$~decays with branching fractions of $10^{-5}$ or larger to obtain a sufficiently large amount of signals in the current data set with a relatively straightforward reconstruction. We use the following kinematic variables to distinguish the signal from the dominant background from \mbox{$e^+e^- \to q\bar{q}$}~continuum events, where $q$ indicates any quark of the first or second generation:
\begin{itemize}
    \item the energy difference $\Delta E \equiv E^{*}_{B} - \sqrt{s}/2$ between the energy $E^{*}$ of the reconstructed $B$ candidate and half of the collision energy~$\sqrt{s}$, both measured in the $\Upsilon(4S)$ frame;
    \item the beam-energy-constrained mass $M_{\rm bc} \equiv \sqrt{s/(4c^4) - (p^{*}_B/c)^2}$, which is the invariant mass of the $B$ candidate where the $B$ energy is replaced by half the collision energy, which is more precisely known. 
\end{itemize}

The signal reconstruction procedure, the event selection criteria and the training of the flavor tagger  are \mbox{developed} and finalized using Monte~Carlo~(MC)~simulation prior to applying it to the experimental data. Experimental and simulated data are then compared in terms of signal yields, background levels, wrong-tag fractions, tagging efficiencies and relevant distributions.

\section{The Belle II detector}
Belle II is a particle-physics detector~\cite{Kou:2018nap, Abe:2010sj}, designed to reconstruct the products of electron-positron collisions produced by the SuperKEKB asymmetric-energy collider~\cite{Akai:2018mbz}, located at the KEK laboratory in Tsukuba, Japan. Belle II comprises several subdetectors arranged around the interaction space-point in a cylindrical geometry. The innermost subdetector is the vertex detector, which uses position-sensitive silicon layers to sample the trajectories of charged particles (tracks) in the vicinity of the interaction region to extrapolate the decay positions of their long-lived parent particles. The vertex detector includes two inner layers of silicon pixel sensors and four outer layers of silicon microstrip sensors. The second pixel layer is currently incomplete and covers only a small portion of azimuthal angle. Charged-particle momenta and charges are measured by a large-radius, helium-ethane, small-cell central drift chamber, which also offers charged-particle-identification information through a measurement of particles' energy-loss by specific ionization. A Cherenkov-light angle and time-of-propagation detector surrounding the chamber provides charged-particle identification in the central detector volume, supplemented by proximity-focusing, aerogel, ring-imaging Cherenkov detectors in the forward regions. A CsI(Tl)-crystal electromagnetic calorimeter allows for energy measurements of electrons and photons.  A solenoid surrounding the calorimeter generates a uniform axial 1.5\,T magnetic field filling its inner volume. Layers of plastic scintillator and resistive-plate chambers, interspersed between the
magnetic flux-return iron plates, allow for identification of $K^0_{\rm L}$ and muons.
The subdetectors most relevant for this work are the silicon vertex detector, the tracking drift chamber, the particle-identification detectors, and the electromagnetic calorimeter.

\section{Selection and reconstruction of signal $B$~candidates}
\label{sec:selection}

We reconstruct the following signal \PB~decays (charge-conjugate processes are implied everywhere),
\vspace{0.5cm}

\begin{tabular}{r l r l}
$\bullet$ & $\PBplus \to \APDzero\, \Pgpp$,  &
\hspace{1.2cm} $\bullet$ & $\PBzero \to \PDminus\, \Pgpp$,\\
$\bullet$ & $\PBplus \to \APDzero\, \Prhoplus$, &
\hspace{1.2cm} $\bullet$ & $\PBzero \to \PDminus\, \Prhoplus$, \\
$\bullet$ & $\PBplus \to \APD^{*0}(\to\APDzero\,\Pgpz)\, \Pgpp$,  &
\hspace{1.2cm} $\bullet$ & $\PBzero \to \PD^{*-}(\to\APDzero\,\Pgpm)\, \Pgpp$,\\
$\bullet$ & $\PBplus \to \APD^{*0}(\to\APDzero\,\Pgpz)\, \Prhoplus$, &
\hspace{1.2cm} $\bullet$ & $\PBzero \to \PD^{*-}(\to\APDzero\,\Pgpm)\, \Prhoplus$, \\
$\bullet$ & $\PBplus \to \APD^{*0}(\to\APDzero\,\Pgpz)\,  \Pa_1^+$, &
\hspace{1.2cm} $\bullet$ & $\PBzero \to \PD^{*-}(\to\APDzero\,\Pgpm)\, \Pa_1^+$, \\
\end{tabular}\vspace{0.5cm}

for which we reconstruct the following \PD~decays,
\vspace{0.5cm}

\begin{tabular}{r l r l}
$\bullet$ & $\APDzero\to\PKplus\Pgpm$,  &
\hspace{1.2cm} $\bullet$ & $\PDminus\to\PKp\Pgpm\Pgpm$,\\
$\bullet$ & $\APDzero\to\PKplus\Pgpm\Pgpp\Pgpm$, &
\hspace{1.2cm} $\bullet$ & $\PDminus\to\PKzS\,\Pgpm$, \\
$\bullet$ & $\APDzero\to\PKplus\Pgpm\Pgpz$, &
\hspace{1.2cm} $\bullet$ & $\PDminus\to\PKzS\,\Pgpm\Pgpz$,\\
$\bullet$ & $\APDzero\to\PKzS\,\Pgpp\Pgpm$, &
\hspace{1.2cm} $\bullet$ & $\PDminus\to\PKp\Pgpm\Pgpm\Pgpz$. \\
\end{tabular}    

%%%%%%%%%% Data/tools %%%%%%%%%%%%

\subsection{Data}
We use generic MC~simulation to optimize the event selection and compare the flavor distributions and fit results obtained from the experimental data with expectations.
The generic MC~simulation consists of samples that include $\PBzero\APBzero$, $\PBplus\PBminus$, $\Pqu\Paqu$, $\Pqd\Paqd$, $\Pqc\Paqc$, and $\Pqs\APqs$~processes in proportions representing their different production cross~sections and correspond to an integrated luminosity of 50\,fb$^{-1}$, about six times the $\Upsilon$(4S) data. In addition, we generate $2\cdot 10^7$ signal-only events~\cite{Ryd:2005zz}, where the signal $\PB$~meson decays to the invisible final state $\PBzero\to\Pgngt\Pagngt$ and the tag-side $\PB$~meson decays to any possible final state according to the known branching fractions. \newpage

As for experimental data, we use all 2019  $\Upsilon$(4S) good-quality runs, corresponding to an integrated luminosity of $8.7\pm 0.2\,\si{fb^{-1}}$~\cite{Abudin_n_2020}. All events are required to meet loose data-skim selection criteria, based on total energy and charged-particle multiplicity in the event, targeted at reducing sample sizes to a manageable level. All data are processed using the Belle~II analysis software framework~\cite{Kuhr:2018lps}.

\subsection{Reconstruction and baseline selection}
%We form final-state particle candidates by  applying loose baseline selection criteria and then combine them according to the topologies of the desired decays to reconstruct intermediate states and $B$ candidates. \par 

We reconstruct charged pion and kaon candidates by starting from the most inclusive charged-particle classes and by requiring fiducial criteria that restrict them to the full acceptance in the central drift chamber and to loose ranges in impact parameter to reduce beam-background-induced tracks, which do not originate from the interaction region. Additionally, we use charged-particle identification information to identify kaon candidates.
We reconstruct neutral pion candidates by requiring photons to exceed energies of about $30$\,MeV, restricting the diphoton mass to be in the range \mbox{$\SI{120} <\; M(\gamma\gamma) < \SI{145}{MeV}/c^2$}. The mass of the $\pi^0$ candidates is constrained to its known value in subsequent kinematic fits. 
For $K_{\rm S}^0$ reconstruction, we use pairs of oppositely charged particles that originate from a common point in space or vertex position and have a dipion mass in the range \mbox{$\SI{450} <\; M(\pi^+\pi^-) < \SI{550}{MeV}/c^2$}. 
% To reduce combinatorial background, we apply, depending on the $K_{\rm S}^0$ momentum, additional requirements on the distance between the two charged-pion candidates, the angle between the direction of the pion-pair momentum and the direction of the common space-point from the collision point, and the distance between the common space-point and the collision point. 
To reduce combinatorial background, we apply additional requirements, dependent on $K_{\rm S}^0$~momentum, on the distance between trajectories of the two charged-pion candidates, the $K^0_{\rm S}$~flight distance, and the angle between the pion-pair momentum and the direction of the $K^0_{\rm S}$~flight.

The resulting $K^\pm$, $\pi^\pm$, $\pi^0$, and $\PKzS$ candidates are combined to form  $\PD^{(*)}$~candidates in the various final states, by requiring their invariant masses to satisfy:
\begin{itemize}
\item $\SI{1.84} <\; M(\PKp\Pgpm,\, \PKp\Pgpm\Pgpp\Pgpm,\, \PKp\Pgpm\Pgpz,\, \PKzS\,\Pgpp\Pgpm) < \SI{1.89}{GeV}/c^2$,

\item $\SI{1.844} <\; M(\PKp\Pgpm\Pgpm,\, \PKzS\,\Pgpm,\, \PKzS\,\Pgpm\Pgpz,\, \PKp\Pgpm\Pgpm\Pgpz) < \SI{1.894}{GeV}/c^2$,

\item $\SI{0.14} <\; M(\PDzero\Pgpz) - M(\PDzero) < \SI{0.144}{GeV}/c^2$,

\item $\SI{0.143} <\; M(\PDzero\Pgpp) - M(\PDzero)< \SI{0.147}{GeV}/c^2$.
\end{itemize}
 We reconstruct $\Prhopm$ candidates from pairs of charged and neutral pions, and $\Pa_1^{\pm}$ candidates from three charged pions, by requiring the following  conditions:
\begin{itemize}
    \item $ \vert  M(\Pgpp\Pgpz) - M_{\Prho} \vert < \SI{0.1}{GeV}/c^2$,
\item  $ \vert M(\Pgpp\Pgpm\Pgpp) - M_{\Pa_1}\vert < \SI{0.4}{GeV}/c^2$,
\end{itemize}
where $M_{\Prho}$ and $M_{\Pa_1}$ are the known PDG masses of the $\Prho$~and $\Pa_1$~mesons.
To identify primary~$\Pgppm$~(direct \PB~daughters) and $\Pgppm$ candidates used to reconstruct $\Prhopm$ and $\Pa_1^{\pm}$~candidates, we additionally use charge-particle identification information and require the $\Pgppm$~momentum in the $\PUpsilonFourS$ frame to be larger than $\SI{0.2}{GeV}/c$.

To finalize the reconstruction of signal $\PB$~candidates, we associate the $\PD^{(*)}$ candidates with appropriate additional candidate
particles $\Pgppm$, $\Prhopm$ or $\Pa_1^{\pm}$.  We keep only $\PB$~candidates that fulfill~$M_{\rm bc} > \SI{5.27}{GeV}/c^2$ and $\vert \Delta E\vert < \SI{0.12}{GeV}$. Additionally, for channels with \Prhopm candidates, we remove combinatorial background from soft $\Pgpz$ collinear with the \Prhopm, by requiring the cosine of the helicity angle $\theta_{\rm H}$ between the \PB and the \Pgpp momenta in the \Prho frame to be\linebreak \mbox{ $\cos{\theta_{\rm H}}< 0.8$}. 

We form the tag side of the signal $\PB$~candidates,  using all remaining tracks and photons that fulfill loose fiducial criteria, and KLM clusters.

 %For KLM clusters, we have no additional requirements.
%Since we perform a time-integrated validation in this note, we do not use vertex information. We therefore do not reconstruct the vertices of both signal and tag-side \PB~mesons. 

%To tag the flavor of the tag-side $\PB$~mesons, we use both category-based and DNN flavor taggers. The two algorithms run independently of each other.

\subsection{Continuum suppression and final selection}

To suppress continuum background from light $\Pq\Paq$~pairs, we apply requirements on the two topological variables with the highest discrimination power between signal from hadronic \PB~decays and continuum background: $\cos{\theta_{\rm T}^{\rm sig, tag}}$, the cosine of the angle between the thrust axis of the signal~$\PB$~(reconstructed) and the thrust axis of the tag-side $B$~(remaining tracks and clusters), and $R_2$, the ratio between the second and zeroth Fox-Wolfram moments using the full event information.

We vary the selection on  $\cos{\theta_{\rm T}^{\rm sig, tag}}$ and $R_2$ to maximize the figure of merit ${\rm S}/\sqrt{{\rm S}+{\rm B}}$, where ${\rm S}$ and ${\rm B}$ are the number of signal and background \PB~candidates in the range \mbox{$M_{\rm bc}>5.27\,\si{GeV}/c^2$} and \mbox{$\SI{-0.12} < \Delta E < \SI{0.09}{GeV}$}.
%We determine signal and background yields using MC~information. 
Both $\cos{\theta_{\rm T}^{\rm sig, tag}}$ and $R_2$ requirements are optimized simultaneously using simulation.  %For the optimization we select randomly one candidate for each channel but allow for multiple candidates from different channels. In this way, we reduce the influence of the channels with low purity on the result of the optimization. 
We optimize the requirements for charged and for neutral candidates independently. The optimized requirements are found to be $\cos{\theta_{\rm T}^{\rm sig, tag}} < 0.87$ and $ R_2 < 0.43$ for charged $\PB$~candidates, and $\cos{\theta_{\rm T}^{\rm sig, tag}} < 0.95$ and $ R_2 < 0.35$ for neutral $\PB$~candidates. 

After applying the $\cos{\theta_{\rm T}^{\rm sig, tag}}$ and $R_2$ requirements, more than one candidate per event populates the resulting $\Delta E$~distributions, with average multiplicities for the different channels ranging from 1.00 to 7.89 (about $75\%$ of the channels have multiplicities between 1.00 and 3.00). We select a single $\PB$~candidate per event randomly to avoid possible bias using a reproducible pseudo-random ranking.  
%First, we select one random candidate per event for each channel separately, and then we select only one random candidate among the different channels. 
The analyses of charged and neutral $\PB$~channels are independent: we select one random candidate among the charged and one among the neutral channels independently.
% The main challenge in reconstructing significant charmless signals is the large contamination from continuum background. To discriminate against such background, we use a binary boosted decision-tree classifier that combines nonlinearly a number of variables known to provide statistical discrimination between $B$-meson signals and continuum.   We choose 39 variables whose correlation with $\Delta E$ and $M_{\it bc}$ is below $\pm$5\% to avoid biases in signal-yield determination. These variables include quantities associated to event topology (global and signal-only angular configurations), flavor-tagger information, vertex separation and uncertainty information, and kinematic-fit quality information. 
% Data-simulation comparison for input distributions using the control sample shows no major inconsistency for both signal and background.
% We train the classifier to identify statistically significant signal and background features using unbiased simulated samples. %Overtraining is explicitly tested and ruled out.

%%% Tagging algorithm %%%% 

\section{The tagging algorithm} 

\label{sec:algos}

 We determine the flavor of the tag side using the Belle~II category-based flavor tagger~\cite{Abudinen:2018}. The category-based flavor tagger is a multivariate algorithm that receives as input kinematic and PID information of the particles in the tag side, and provides as output the product $q\cdot r$, where $q$ is the flavor of the tag-side $\PB$~meson, and $r$ the dilution factor. A dilution factor $r=0$ corresponds to a
fully diluted flavor (no possible distinction between $\PBzero$ and $\APBzero$) and a dilution factor $r=1$ to a perfectly tagged flavor. By convention $q = +1$ corresponds to a tag-side \PBzero, and $q=-1$  to a tag-side \APBzero.

The algorithm relies on flavor-specific decay modes. Each decay mode has a particular decay topology and provides a flavor specific signature. Similar or complementary decay modes are combined to obtain additional flavor signatures. The different flavor signatures are sorted into thirteen tagging categories. Table~\ref{table:targets} shows an overview of all thirteen categories together with the underlying decay modes. 

\begin{figure}[tb]
\centering
\includegraphics[width=0.98\textwidth]{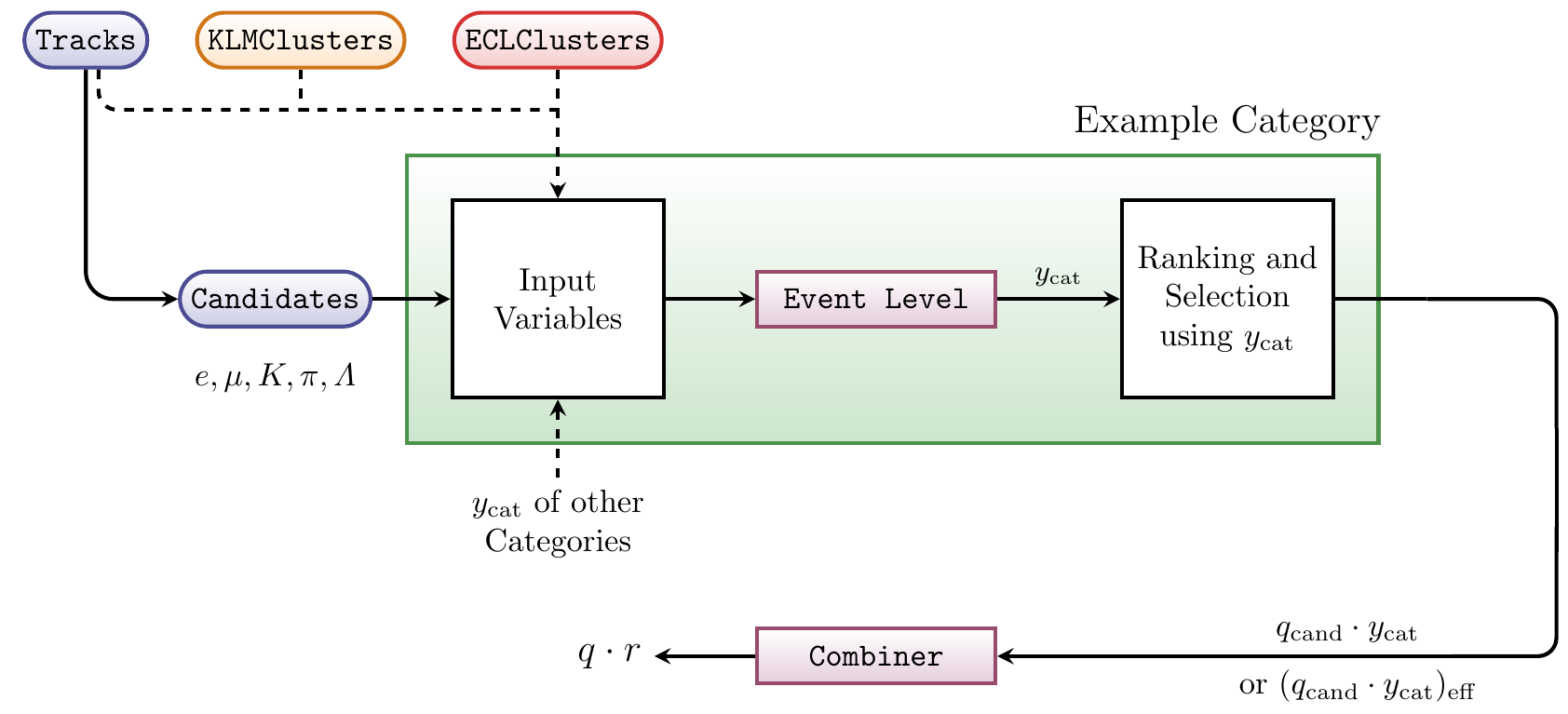}
\caption{Procedure for each single category (green box): the candidates correspond to the reconstructed tracks for a specific mass hypothesis. Some of the input variables consider all reconstructed tracks and all neutral ECL and KLM clusters on the tag side. The magenta boxes represent multivariate methods: $y_{\text{cat}}$ is the output of the event level. The output of the combiner is equivalent to the product $q\cdot r$.}
%\caption{Procedure steps of the Flavor Tagger: The grey path on top illustrates only the steps of the procedure. Reconstructed tracks are available for five different mass hypotheses. Each green box corresponds to a multivariate method: $y_{\text{track}}$ and $y_{\text{cat}}$ correspond to the output of track and event levels respectively.  The variables $(q\cdot y_{\text{cat}})_w$ represent the weighted sum over the three tracks with highest $y_{\text{cat}}$.}
  \label{fig:FlavTagSingleCat}
\end{figure}

The algorithm performs a two-level procedure with an event level for each category followed by a combiner level. Figure~\ref{fig:FlavTagSingleCat} illustrates the procedure. At the event~level, the algorithm identifies decay products providing flavor signatures among the $\Pepm$, $\Pmupm$, $\PKpm$, $\Pgppm$ and $\PLambda$ candidates in the tag side using Fast Boosted Decision Tree~(FBDT)~\cite{Tkeck} classifiers. At the combiner level, the algorithm combines the information provided by all categories into the final product $q\cdot r$ using a combiner-level FBDT. This classifier receives an input from each category corresponding to the product $q_{\rm cand}\cdot y_{\rm cat}$, where $q_{\rm cand}$ is the charge of the candidate identified as flavor-specific decay product, and $y_{\rm cat}$ is the probability provided by the event-level FBDT. Only for the Kaon and the Lambda category, the input is the effective product $(q_{\rm cand}\cdot y_{\rm cat})_{\rm eff}$ of the three candidates with the highest probability. 

The algorithm is trained using signal MC~events where the signal $\PB$~meson decays to the invisible final state $\PBzero\to\Pgngt\Pagngt$. Using the $\PBzero\to\Pgngt\Pagngt$ samples, we avoid possible bias due to \CP~asymmetries or reconstruction performance since these samples are generated without built-in \CP violation,  and all reconstructed objects~(tracks, photons and KLM clusters) can be used to form the tag side without passing through reconstruction of the signal side. The flavor tagger is trained with a sample of about $10^7$ MC~events and tested with an independent sample of the same size to exclude overtraining.

{\renewcommand{\arraystretch}{1.1}
\begin{table}[htb]
\caption{Tagging categories and their targets~(left) with examples of the considered decay modes~(right). Here, $p^*$ stands for momentum in the center-of-mass frame and~$\Plpm$ for charged leptons (\Pmuon or \Pelectron). }
\begin{center}
    \label{table:targets} \vspace{0.4cm}
  \begin{minipage}[ht]{0.55\linewidth}
  \begin{tabular}{ @{}c@{} @{}l@{} }
    \begin{tabular}{ l  @{}c@{} }
    \hline
    Categories & Targets for \APBzero \\ \hline\hline
    Electron & \color[rgb]{0.000,0.3,0} $\Pelectron$\\ %   
    Intermediate Electron & \color[rgb]{0.8,0.1,0.5}\ $\APelectron$ \\
    Muon &\color[rgb]{0.000,0.3,0} $\Pmuon$  \\
    Intermediate Muon & \color[rgb]{0.8,0.1,0.5}\ $\APmuon$\\
    Kinetic Lepton & \color[rgb]{0.000,0.3,0} $\Pleptonminus$ \\
    Intermediate Kinetic Lepton& \color[rgb]{0.8,0.1,0.5}\ $\Pleptonplus$\\
    Kaon & \color[rgb]{0.8,0,0} $\PKm$ \\
    Kaon-Pion & \thinspace \thinspace {\color[rgb]{0.8,0,0} $\PKm$}, \color[rgb]{0,0.44,0.75} $\Pgpp$ \thinspace \\
    Slow Pion & \color[rgb]{0,0.44,0.75} $\Pgpp$  \\
    Maximum $p^*$ & {\color[rgb]{0.000,0.3,0} $\Plm$}, \color[rgb]{1,0.4,0} $\Pgpm$  \\
    Fast-Slow-Correlated (FSC)  & {\color[rgb]{0.000,0.3,0} $\Pleptonminus$}, \color[rgb]{0,0.44,0.75} $\Pgpp$\\
    Fast Hadron & \color[rgb]{1,0.4,0} $\Pgpm$, $\PKm$ \\
    Lambda & \color[rgb]{0.44,0.18,0.63}$\PLambda$ \\
    \hline
%    \hline\multicolumn{2}{|c|}{Total = 13} \\ \hline
    \end{tabular}
    \end{tabular}
    \end{minipage}
    \begin{minipage}[ht]{0.3\linewidth}
     \vspace{-0.21cm}
     \begin{tabular}{ c }
     Underlying decay modes
     \end{tabular}
     \begin{center}
    \includegraphics{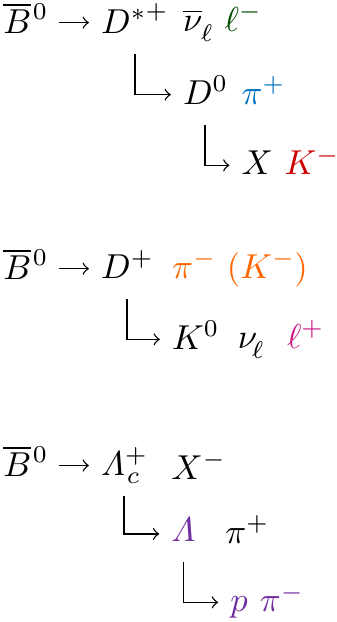}
    \end{center}
     \end{minipage}
\end{center}
\end{table}}

\clearpage

%%% FITS %%%% 

\section{Determination of efficiencies and wrong-tag fractions}
\label{sec:fullFit}

% Signal yields are determined with maximum likelihood fits of the unbinned $\Delta E$ distributions of candidates restricted to the signal region in $M_{\rm bc}$. Fit models are generally determined empirically by using simulation, with the only additional flexibility of a global shift of peak positions when suggested by likelihood-ratio tests. 

The tagging efficiency of the flavor tagger corresponds to the fraction of events to which a flavor tag can be assigned. Since the algorithm needs only one charged track on the tag side to provide a tag, the tagging efficiency is close to $100\%$, with good consistency between data and simulation as Table~\ref{tab:tagging_efficiencies} shows. 

\begin{table}[h!]
    \centering
    \caption{Tagging efficiencies $\varepsilon\pm\delta\varepsilon$ for charged and neutral \mbox{$\PB\to\PD^{(*)}h^{+}$}~candidates in data and in simulation. All values are given in percent. The uncertainties are only statistical.}
    \begin{tabular}{l c c}\hline\hline
     Channel & MC & Data \\ \hline
      $\PBzero\to\PD^{(*)-}h^{+}$    &  $99.78 \pm 0.02$ & $99.78 \pm 0.04$ \\
      $\PBplus\to\APD^{(*)0}h^{+}$   &  $99.81 \pm 0.01$ & $99.72 \pm 0.04$ \\
      \hline
    \end{tabular}
    \label{tab:tagging_efficiencies}
\end{table}{}

To measure the fraction of wrongly tagged events $w$, we sort the events in bins of the dilution factor $r$ provided by the flavor tagger. To compare with our predecessor experiment, we use the binning introduced by Belle~\cite{Bevan:2014iga}. 

Considering \mbox{$\PUpsilonFourS\to\PBzero\APBzero$}~events, the time-integrated probability to observe an event with signal \PB~flavor $q_{\rm sig}\in\{-1,+1\}$ and tag-side \PB~flavor $q_{\rm tag}\in\{-1,+1\}$ in the $i$-th $r$~bin, is given by
\begin{equation}
\mathcal{P}^{i}(q_\text{sig}, q_\text{tag})=\frac{1}{2}\varepsilon_i\bigg[1 -q_\text{sig}\cdot q_\text{tag}\cdot(1-2w_i)\cdot (1-2\cdot\chi_d)\bigg]\text{,}
% \mathcal{P}(q_\text{sig}, q_\text{tag})=\frac{1}{2}\varepsilon\bigg[1-q_\text{tag}\cdot\Delta w 
% -q_\text{sig}\cdot q_\text{tag}\cdot(1-2w)\cdot(1-2\cdot\chi_d)\bigg]\text{,}
 \label{eqn:PQMixFlavT}
 \end{equation}
 where $\chi_d$ is the $\PBzero-\APBzero$ mixing probability, and $w_i$ and $\varepsilon_i$ are the wrong-tag fraction and the partial tagging efficiency in the $i$-th $r$~bin~(7~bins in total). The expression above is obtained assuming that the signal \PB~flavor is correctly identified and that there is no asymmetry in the performance between $\PBzero$ and $\APBzero$ events. We neglect those possible small asymmetries due to the small size of the currently available data sample. 
The current world average for the $\PBzero-\APBzero$~mixing probability is \mbox{$\chi_d = 0.1858 \pm 0.0011$}~\cite{Amhis:2016xyh}.  

Since we need to consider the background to determine the signal $w_i$ and $\varepsilon_i$, we developed a statistical model with a signal and a background component. We determine the signal yield~$N_{\rm sig}$, the background yield~$N_{\rm bkg}$, the partial efficiencies~$\varepsilon_i$ and the wrong-tag fractions~$w_i$ from an extended maximum likelihood fit to the unbinned distributions of~$\Delta E$, $q_{\rm sig}$ and $q_{\rm tag}$. We checked that the $\Delta E$ distribution is statistically independent from those of $q_{\rm sig}$ and $q_{\rm tag}$ with Pearson correlation coefficients below $2\%$. 
%of candidates restricted to the signal region \mbox{$M_{\rm bc} > \SI{5.27}{GeV}/c^2$}. 

In the fit model, the probability density function~(PDF) for each component $j$ is given by
\begin{equation*}
    \mathcal{P}_j(\Delta E, q_\text{sig}, q_\text{tag}) \equiv \mathcal{P}_j(\Delta E)\cdot \mathcal{P}_j(q_\text{sig}, q_\text{tag})\text{.} 
\end{equation*}

We model the signal $\Delta E$ PDF using a Gaussian plus a Crystal Ball function~\cite{Skwarnicki:1986xj} determined empirically using correctly associated signal MC~events, with the additional flexibility of a global shift of peak position and a global scaling factor for the width as suggested by a likelihood-ratio test. The background $\Delta E$ PDF is modeled using an exponential function with a free-to-float exponent.
%determined from a fit of the full model to simulation.

The flavor PDF $\mathcal{P}(q_\text{sig}, q_\text{tag})$ has the same form for signal and background~(Eq.~\ref{eqn:PQMixFlavT}) with independent $\varepsilon_i$,  $w_i$ and $\chi_d$ parameters for signal and background. We fix the background $\chi_d^{\rm bkg}$ parameter to $0$ as we obtain values compatible with $0$ when we let it float. 

The total extended likelihood is given by
\begin{equation*}
\mathcal{L} \enskip \equiv \prod_i \enskip \frac{\ee^{-\sum_j N_j\cdot \varepsilon_i}}{N^i!} \prod_{k=1}^{N^i} \enskip\sum_{j} N_j \cdot \mathcal{P}_j^i(\Delta E^k, q_\text{sig}^k, q_\text{tag}^k)\text{,}
\end{equation*}
where $i$ extends over the $r$~bins, $k$ extends over the events in the bin~$i$, and $j$ over the two components: signal and background.  The PDFs for the different components have no common parameters. $N^i$ denotes the total number of events in the $i$-th $r$~bin. The partial efficiencies~$\varepsilon_i$ are included in the flavor part of $\mathcal{P}_j$.  Since we can fit only to events with flavor information, the sum of all $\varepsilon_i$ must be $1$. We therefore replace the epsilon for the first bin (with lowest $r$) with 
\begin{equation}
    \varepsilon_{1} = 1 - \sum_{i=2}^{7}\varepsilon_{i}\text{,}
    \label{eq:epsilon1}
\end{equation} 
and obtain its uncertainty $\delta\varepsilon_1$ from the width of the residuals of pseudo-experiments. 

\begin{figure}[htb]
 \centering
 \includegraphics[width=0.475\textwidth]{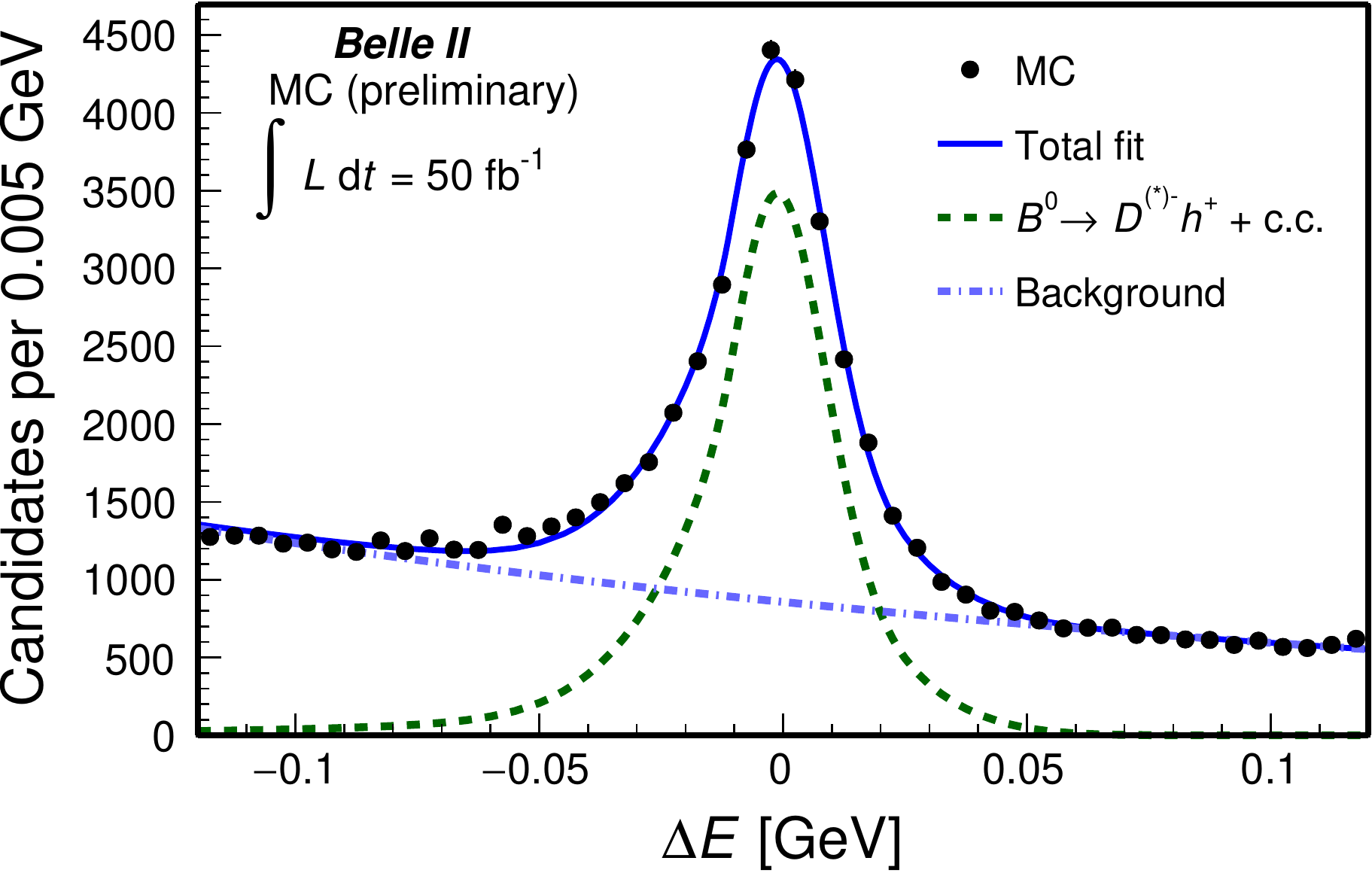}\hfill
 \includegraphics[width=0.475\textwidth]{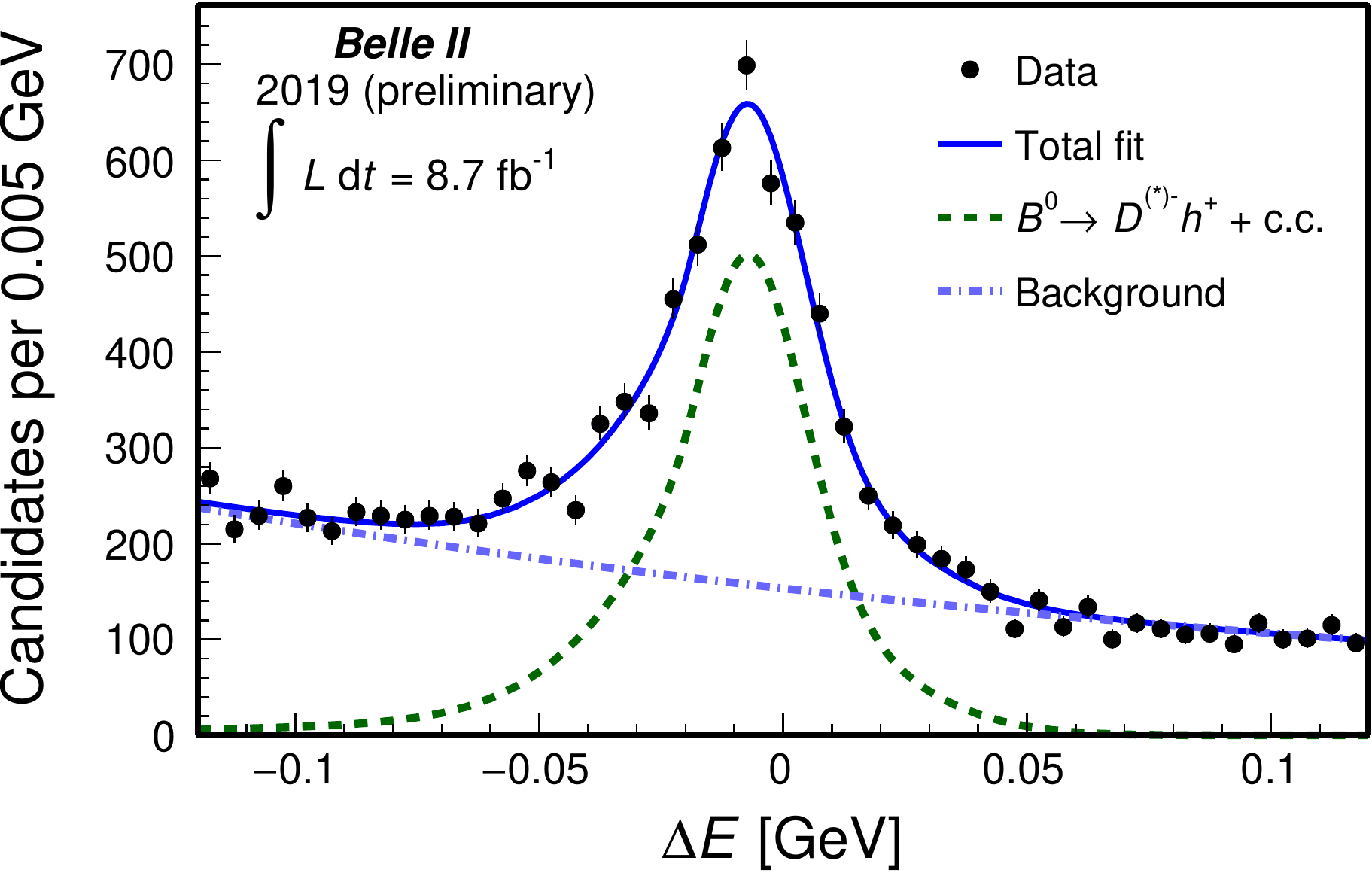}
 \caption{ Fit projection of the maximum likelihood fit to the unbinned distribution of $\Delta E$ for $\PBzero\to\PD^{(*)-}h^{+}$ candidates reconstructed in (left)~simulation and (right)~data, restricted to \mbox{$M_{\rm bc} > 5.27$\,GeV/$c^2$}. The global peak shift and width scaling factor are determined by the fit. }
 \label{fig:fit_dE_neutral_unbinned}
\end{figure}

\begin{figure}[htb]
 \centering
 \includegraphics[width=0.475\textwidth]{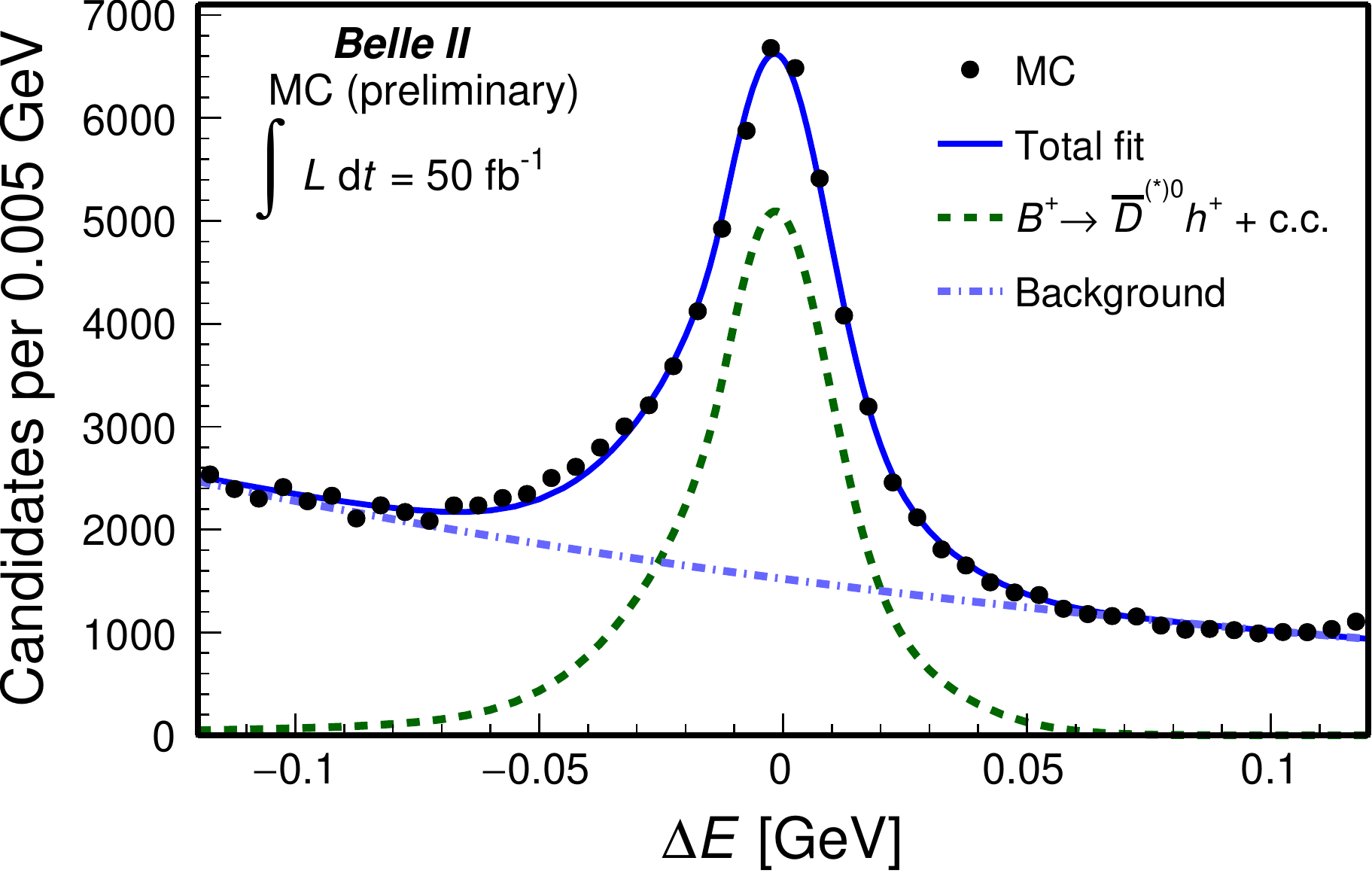}\hfill
 \includegraphics[width=0.475\textwidth]{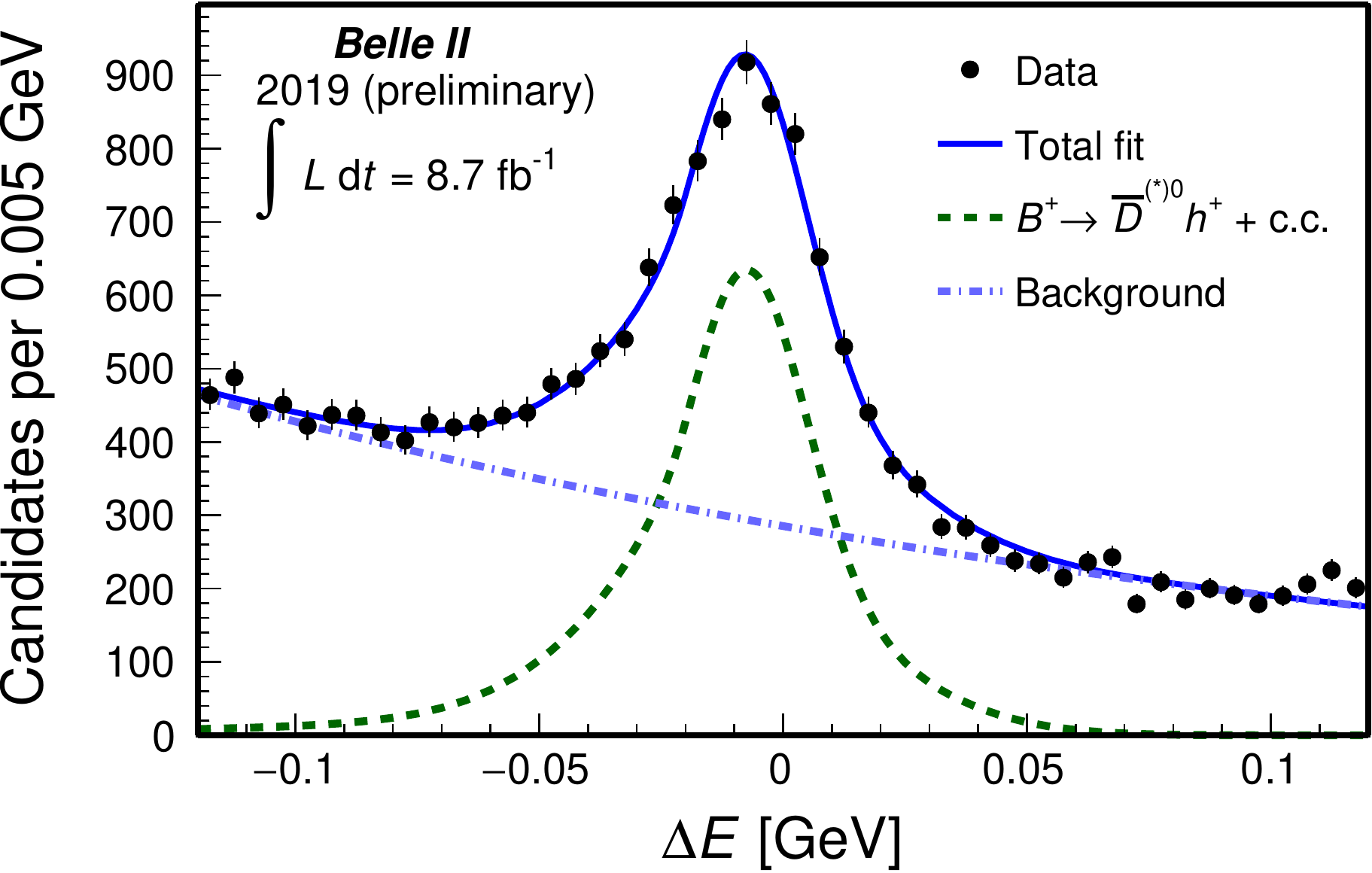}
 \caption{ Fit projection of the maximum likelihood fit to the unbinned distribution of $\Delta E$ for $\PBplus\to\APD^{(*)0}h^{+}$ candidates reconstructed in (left)~simulation and (right)~data, restricted to \mbox{$M_{\rm bc} > 5.27$\,GeV/$c^2$}. The global peak shift and width scaling factor are determined by the fit. }
 \label{fig:fit_dE_charged_unbinned}
\end{figure}

 To validate the $\Delta E$ model, we first perform an extended maximum likelihood fit to the unbinned distribution of $\Delta E$ (without flavor part) in simulation and data.
 %and determine yields $N_{\rm sig}$, $N_{\rm bkg}$, background PDF exponent $c$, global shift $\mu_{\rm C}$ and scaling factor $\sigma_{\rm C}$. 
 Figures~\ref{fig:fit_dE_neutral_unbinned} and \ref{fig:fit_dE_charged_unbinned} show the $\Delta E$ fit projections in data and simulation for charged and neutral $\PB\to\PD^{(*)}h^{+}$ candidates. Table~\ref{tab:yield_summary} summarizes the yields obtained from the fits. We observe a good agreement between data and simulation for neutral \PB~candidates, and lower signal yield with respect to the expectation for charged \PB~candidates. 
 
\newpage

To determine the partial efficiencies~$\varepsilon_i$ and the wrong-tag fractions~$w_i$, we perform a fit of the full model in a single step. 
%There are in total $31$ free parameters. 
For neutral candidates, we additionally leave the signal~$\chi_d^{\rm sig}$ free to float constraining it via a Gaussian constraint,
\begin{equation*}
  \mathcal{L}\; \Rightarrow\; {\rm G}(\chi_d^{\rm sig} - \chi_d,\, \delta\chi_d)\cdot\mathcal{L} \text{,}  
\end{equation*}
where $\chi_d$ and $\delta \chi_d$ are the central value and the uncertainty of the world average. For charged \PB~mesons,  $\chi_d$ is equal to 0 as there is no flavor mixing due to electric charge conservation.

\vspace{1cm}

% \section{Results}

\clearpage

\begin{table}[!ht]
    \centering
    \caption{Summary of yields and yields per integrated luminosity obtained from the fit to MC~simulation and  data. The uncertainties are only statistical.}     
\begin{tabular}{l  r  r  r  r }
\hline\hline
\multicolumn{1}{c}{} & 
\multicolumn{2}{c}{Yield} &   
\multicolumn{2}{c}{Yield/$\si{fb^{-1}}$} \\\hline
$\PBzero\to\PD^{(*)-}h^{+}$ & \multicolumn{1}{c}{MC} & \multicolumn{1}{c}{Data} & \multicolumn{1}{c}{MC} & \multicolumn{1}{c}{Data} \\\hline
   Signal      &	$24246  \pm 251$ &	$4080\pm 114$ &	$485 \pm 5$ &	$ 469 \pm 13$ \\
  Background   &	$43321  \pm 287$ &	$7742\pm 129$ &	$866 \pm 6$ &	$ 890 \pm 15$ \\\hline 
 \\
$\PBplus\to\APD^{(*)0}h^{+}$  & \multicolumn{1}{c}{MC} & \multicolumn{1}{c}{Data} & \multicolumn{1}{c}{MC} & \multicolumn{1}{c}{Data} \\\hline
   Signal      &	$39706 \pm 280$ &	$ 5506 \pm 148$   &	$794  \pm 6$ & 	$ 633 \pm 17$ \\
  Background   &	$77280 \pm 340$ &	$ 14553 \pm 176$  &	$1546 \pm 7$ & 	$ 1673 \pm 20$ \\ 
\hline\hline
\end{tabular}
    \label{tab:yield_summary}
\end{table}

\section{Data/MC comparison for signal and background}

\label{sec:splot}

We check the data/MC agreement of the flavor tagger output in the fit range by performing an $s\mathcal{P}lot$~\cite{Pivk:2004ty} analysis using $\Delta E$ as control variable. We determine $s\mathcal{P}lot$ weights using the fit model developed in Sec.~\ref{sec:fullFit}. We weight the data with the $s\mathcal{P}lot$ weights to obtain the individual signal and background distributions in data and compare them with MC~simulation. We normalize the simulated samples by scaling the total number of events to those observed in data. 

\begin{figure}
\centering
\includegraphics[width=0.7\linewidth]{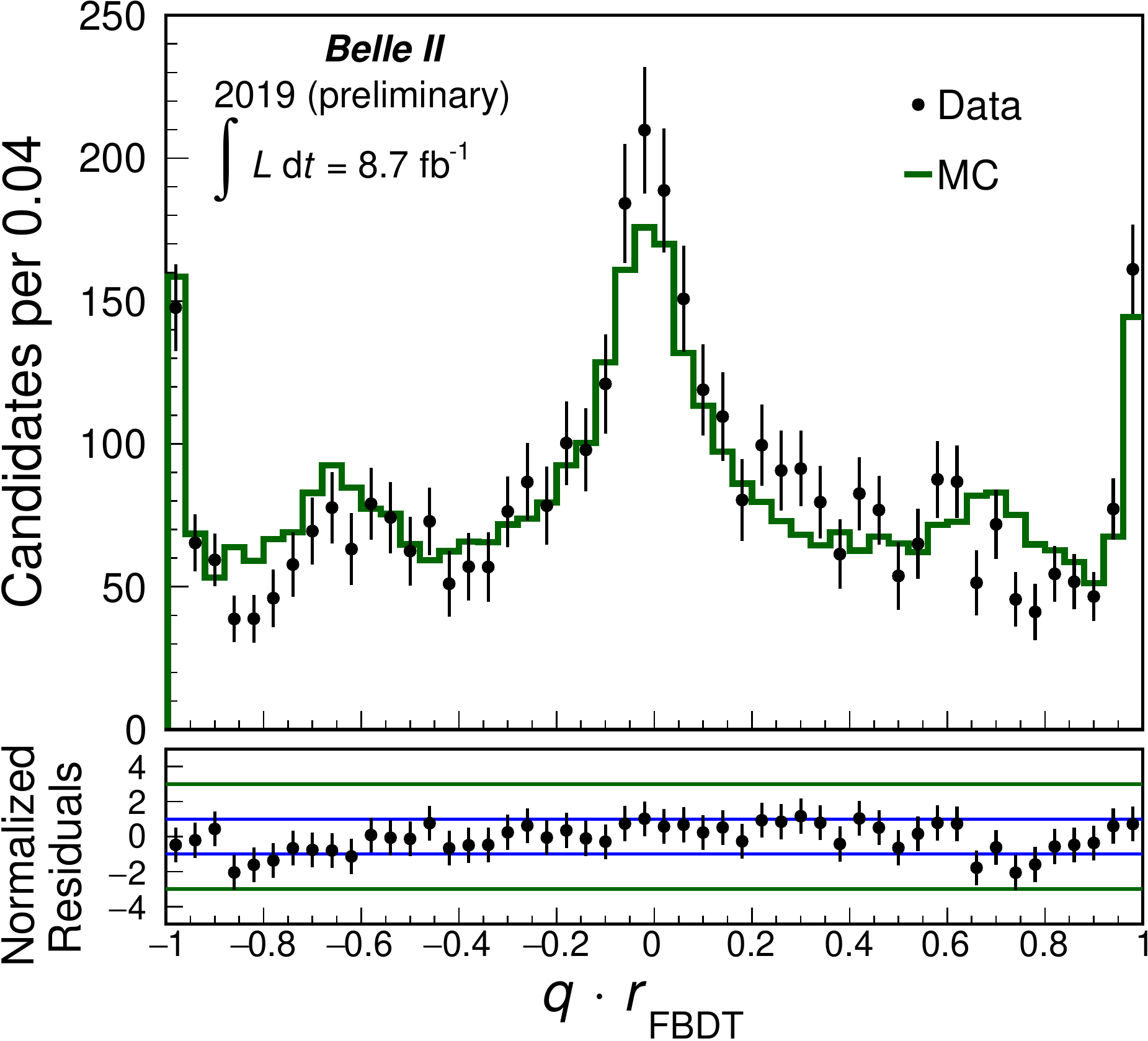}\\
\vspace{1cm}
\includegraphics[width=0.7\linewidth]{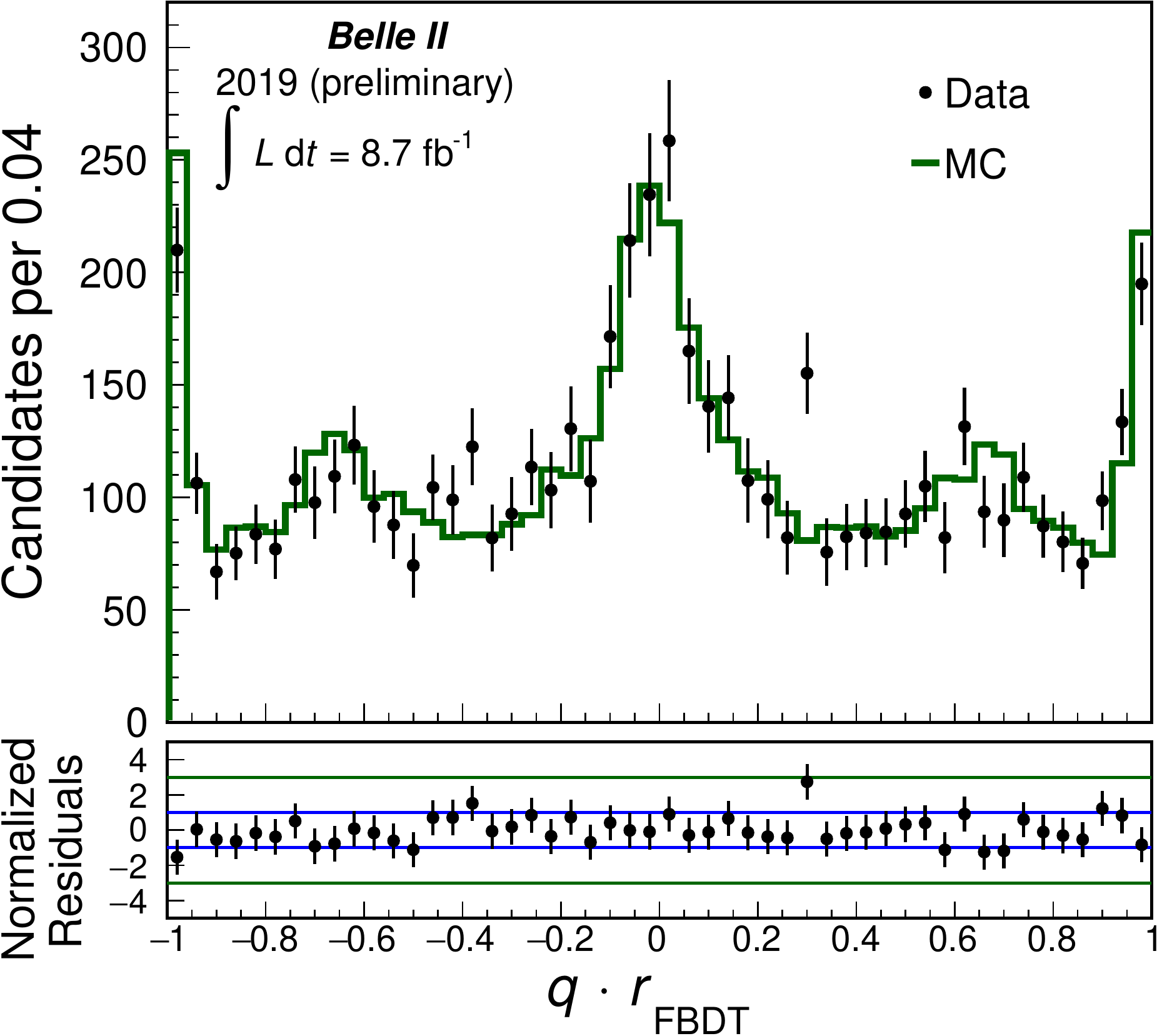}
\caption{\label{fig:qrSigDistData} Normalized $q\cdot r_{\rm FBDT}$ distributions in data and MC~ simulation for (top)~neutral and (bottom)~charged $\PB\to\PD^{(*)}\Ph^{+}$ candidates. The contribution from the signal component in data is compared with correctly associated signal~MC events.}
\end{figure}

\begin{figure}
\centering
\includegraphics[width=0.7\linewidth]{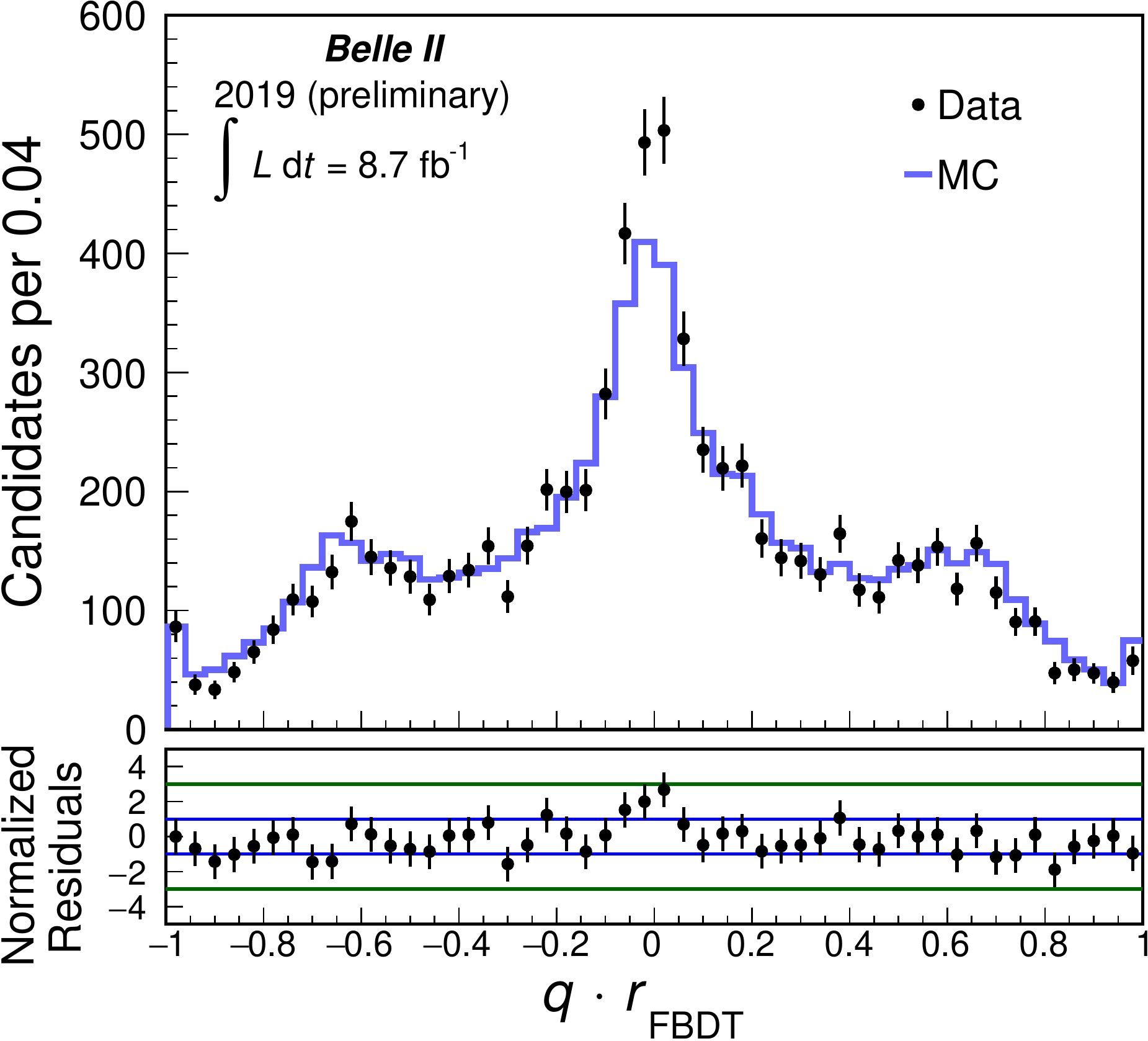}\\
\vspace{1cm}
\includegraphics[width=0.7\linewidth]{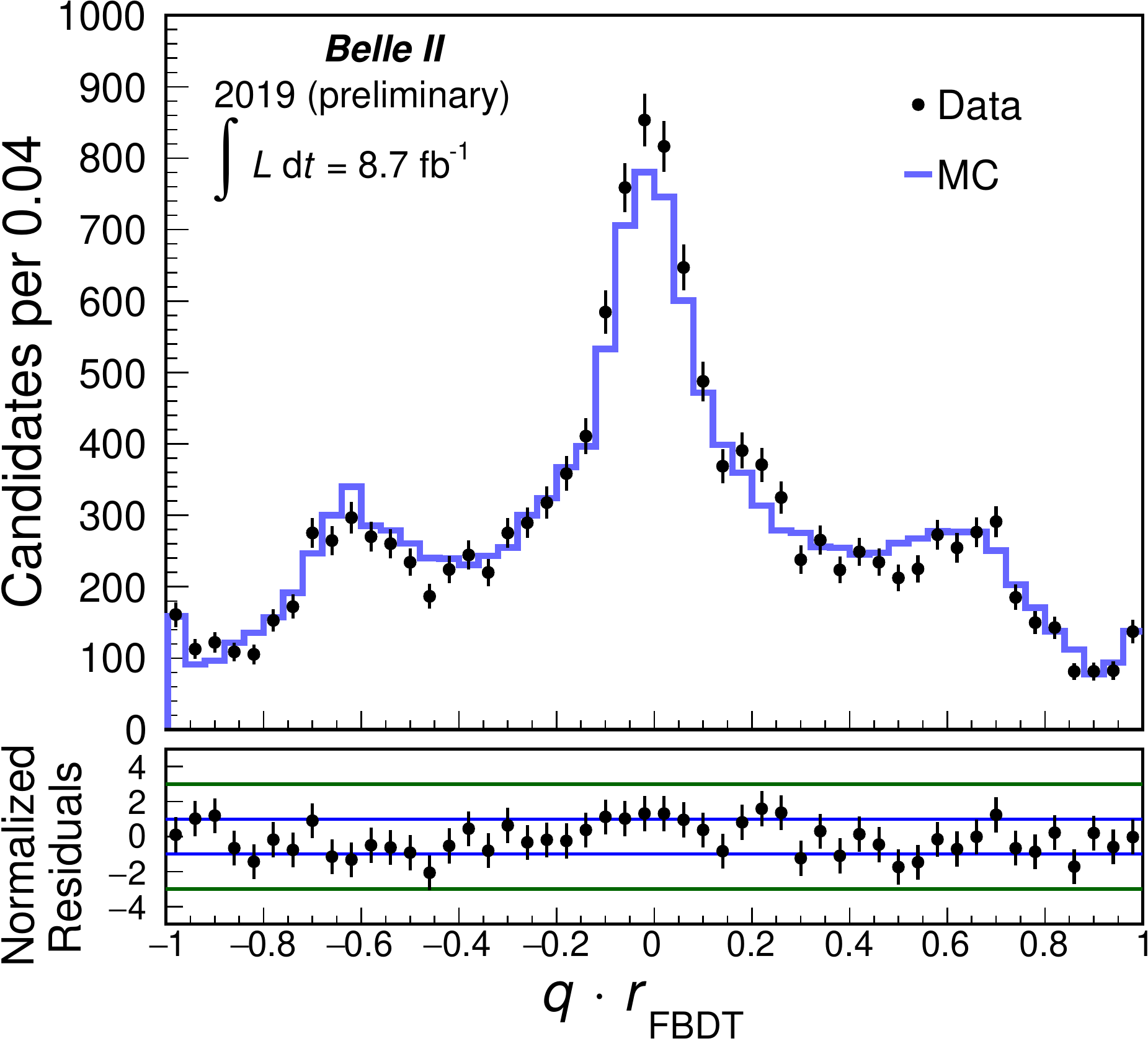}
\caption{\label{fig:qrBkgDistData} Normalized $q\cdot r_{\rm FBDT}$ distributions in data and MC~simulation for (top)~neutral and (bottom)~charged $\PB\to\PD^{(*)}\Ph^{+}$ candidates. The contribution from the background component in data is compared with simulated events in the side band.}
\end{figure}

\begin{figure}
    \centering
    \subfigure{\includegraphics[width=0.46\textwidth]{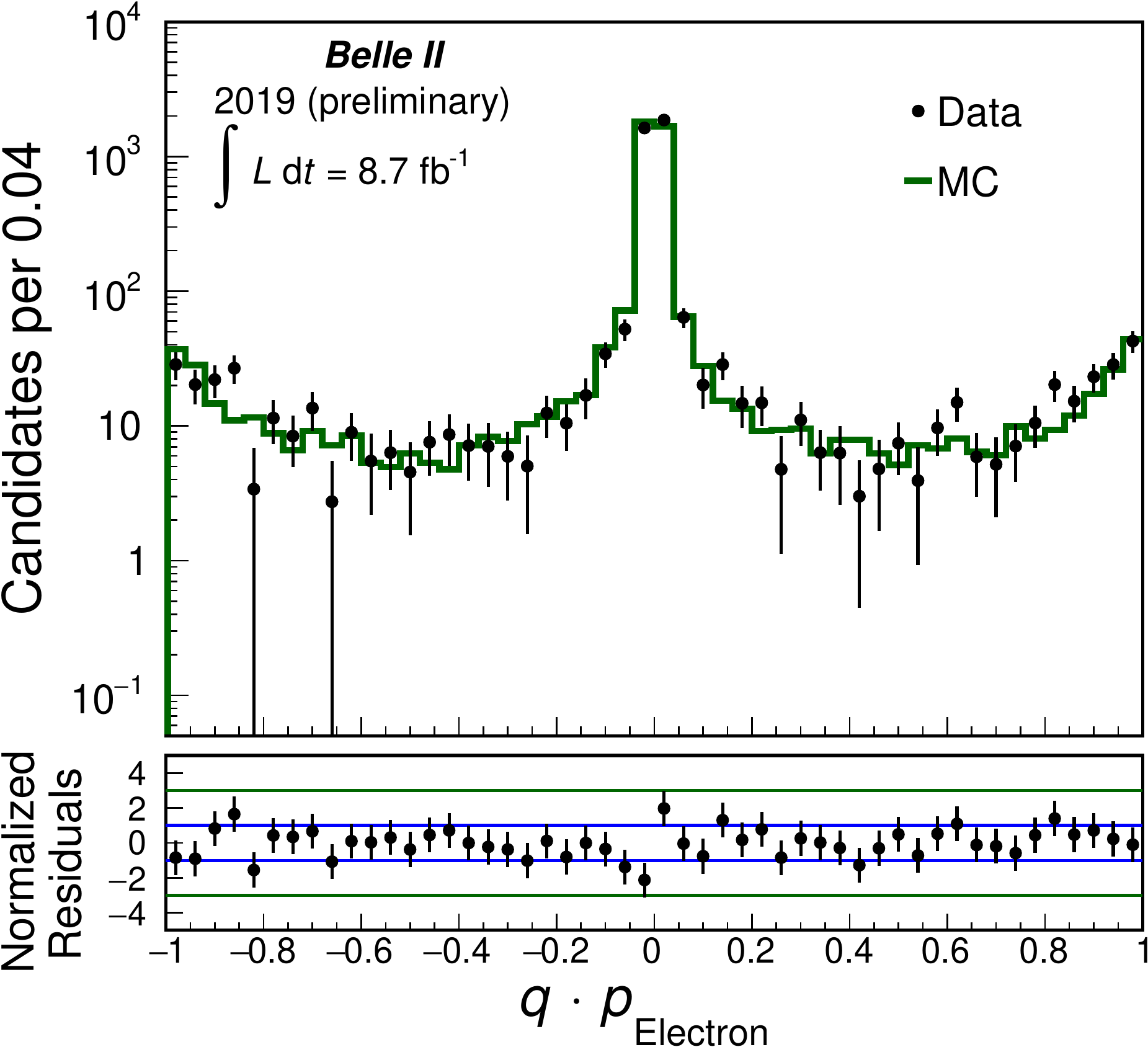}}\hfill 
    \subfigure{\includegraphics[width=0.46\textwidth]{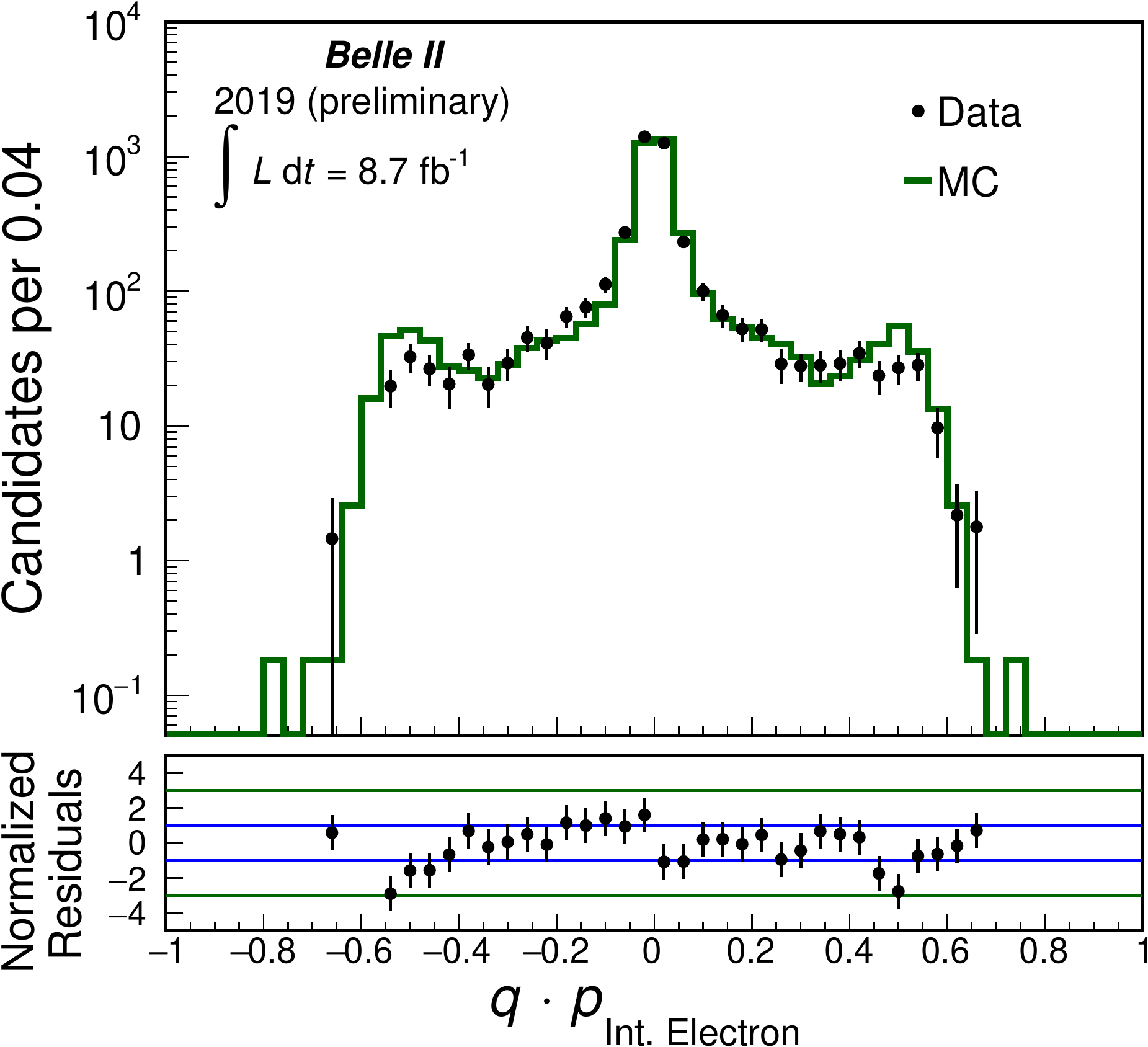}} 
    \subfigure{\includegraphics[width=0.46\textwidth]{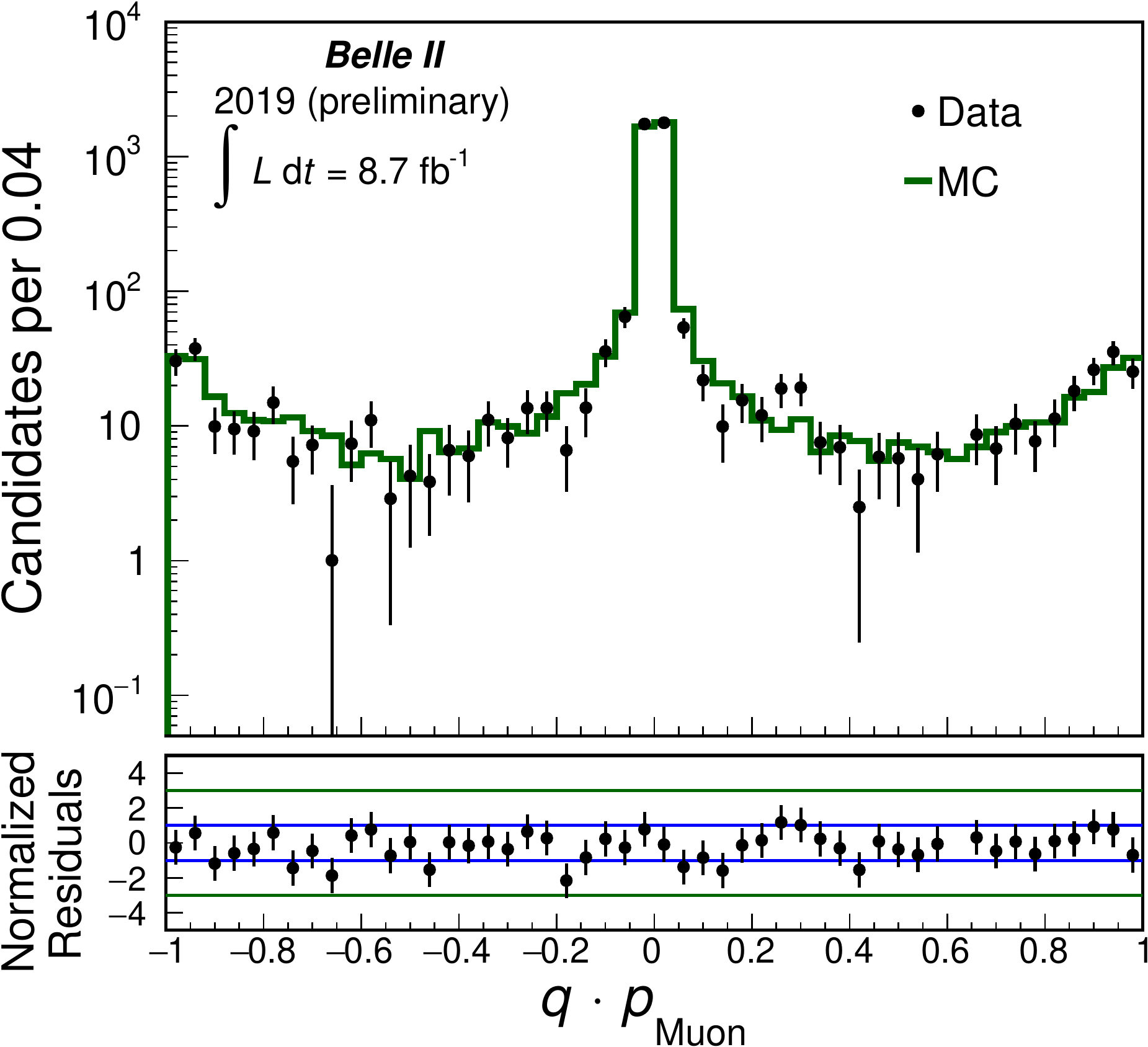}} \hfill 
    \subfigure{\includegraphics[width=0.46\textwidth]{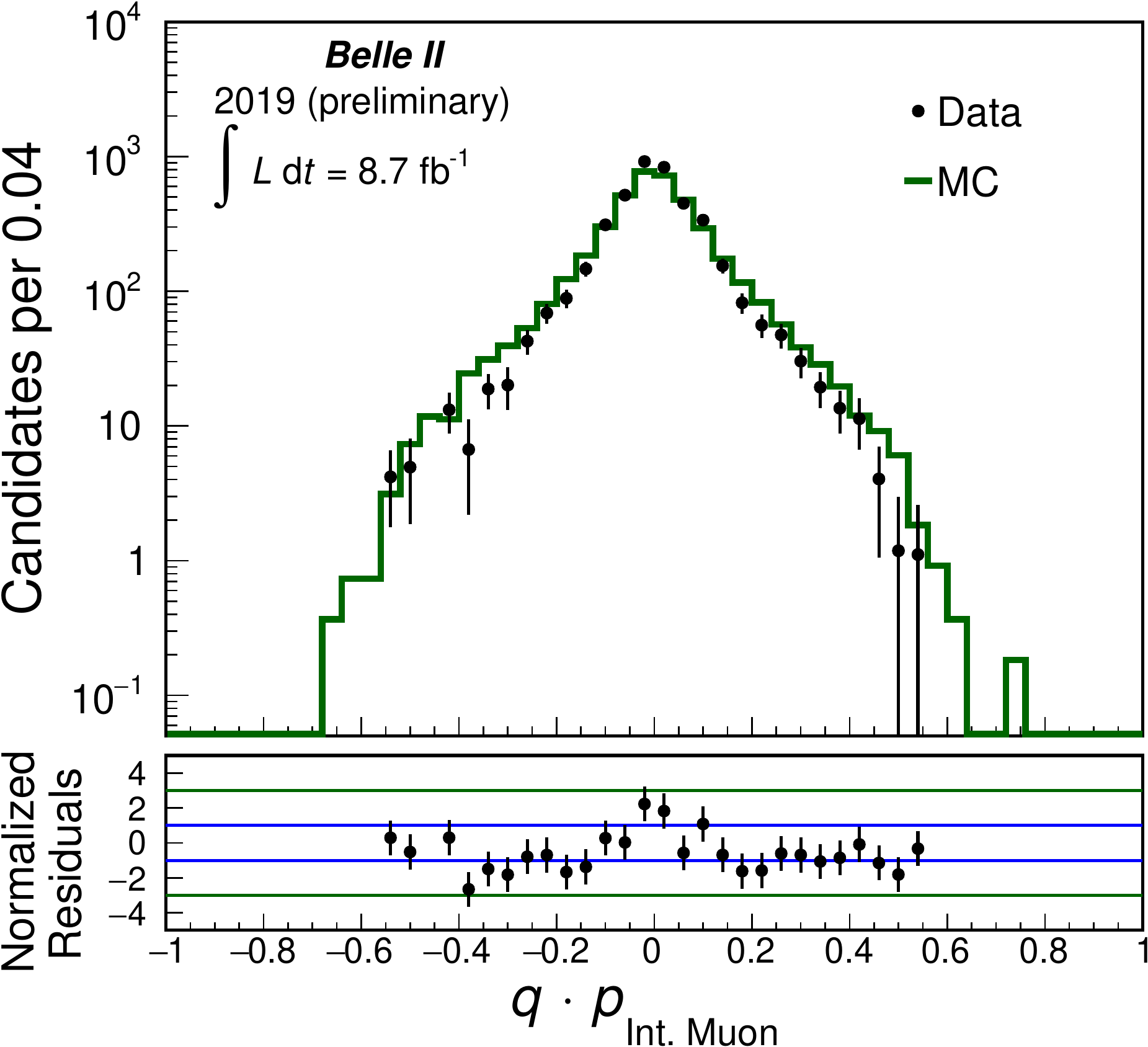}} 
    \subfigure{\includegraphics[width=0.46\textwidth]{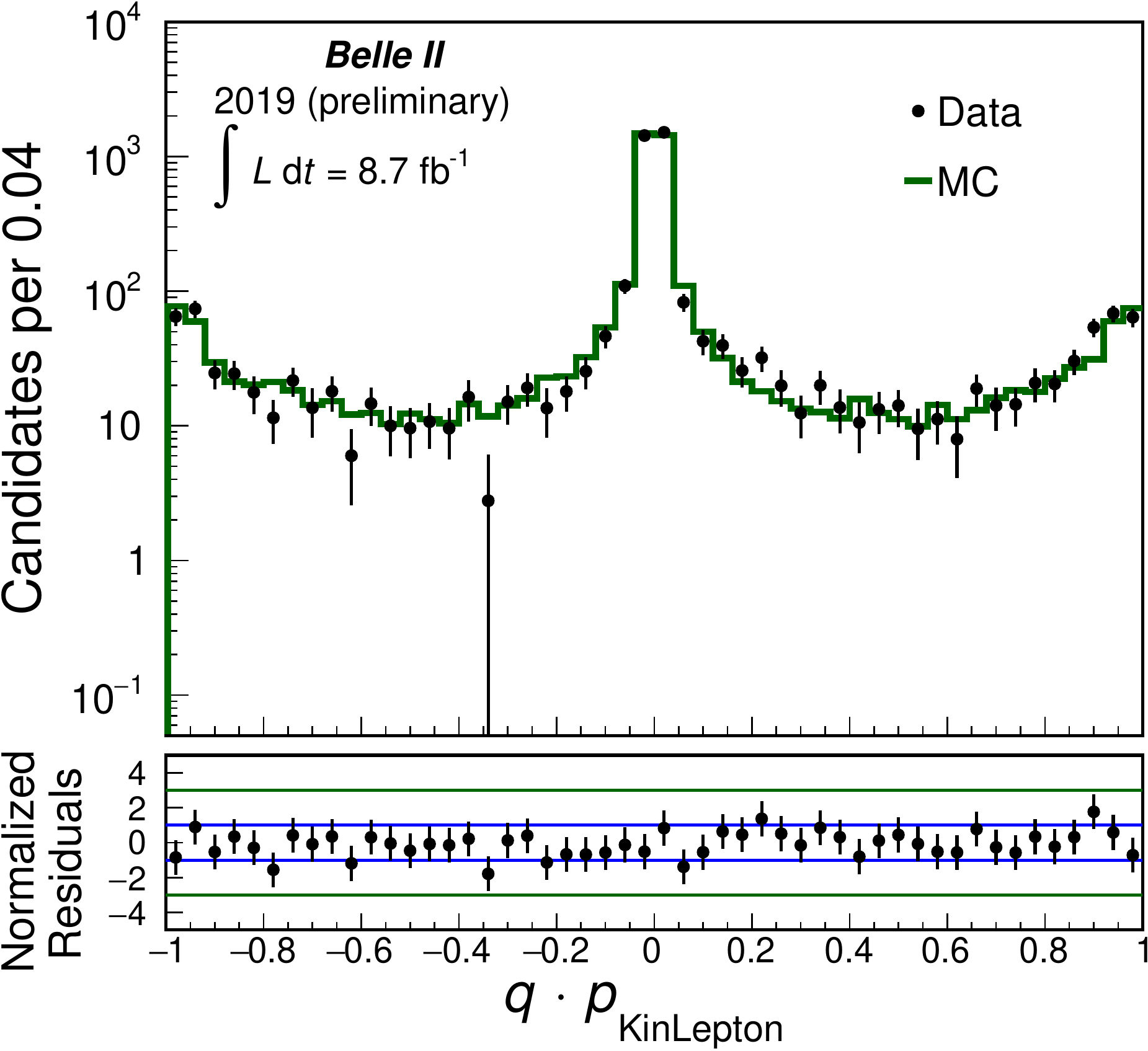}} \hfill
    \subfigure{\includegraphics[width=0.46\textwidth]{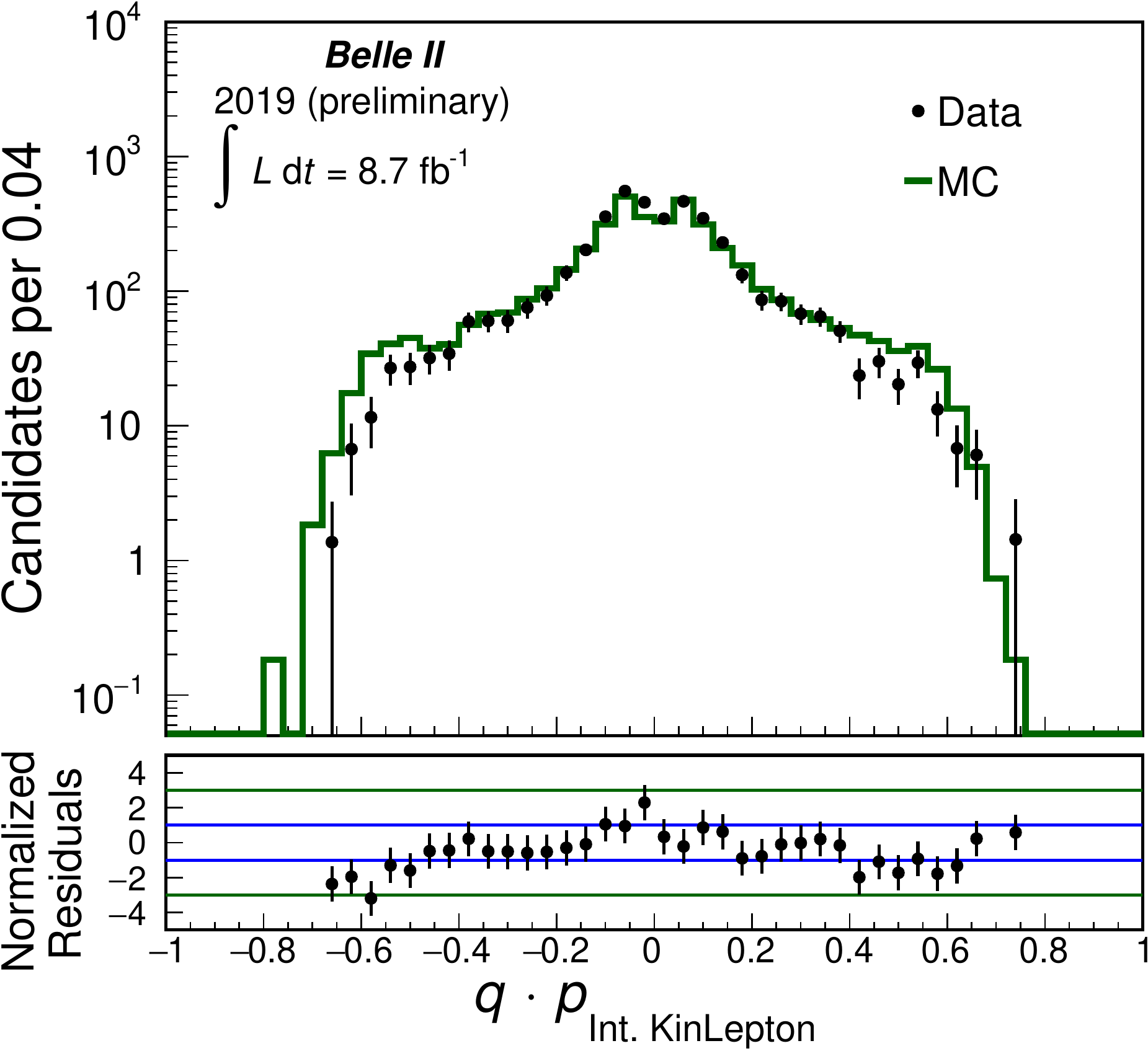}} 
    \caption{
    Normalized output distributions of individual tagging categories  in data and MC~\mbox{simulation} for $\PBzero\to\PD^{(*)-}\Ph^{+}$ candidates. The contribution from the signal component in data is compared with correctly associated signal MC~events~(1/3).}
    \label{fig:QP_data_splotMC_1}
\end{figure}

\begin{figure}
    \centering
    \subfigure{\includegraphics[width=0.46\textwidth]{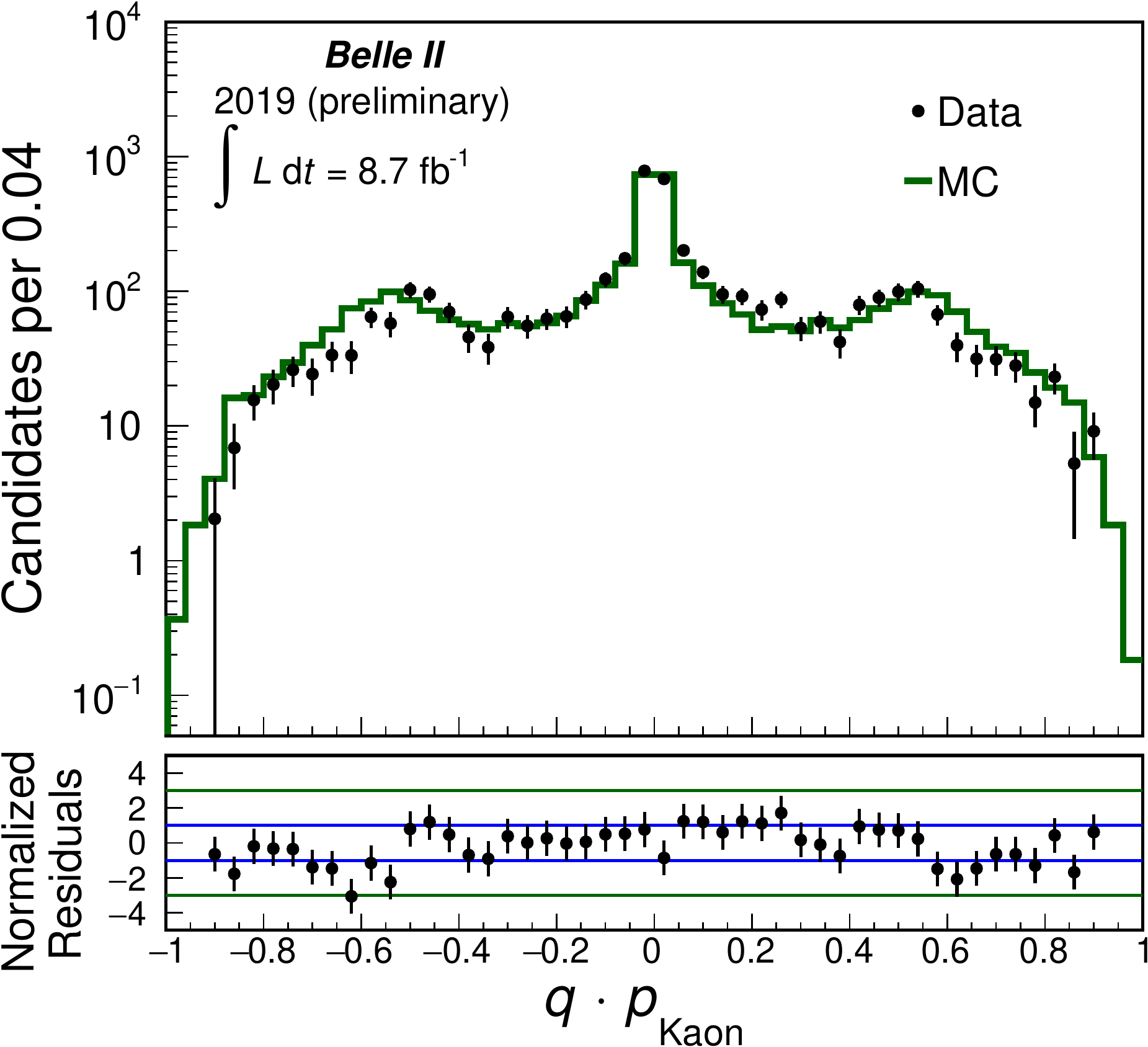}}\hfill 
    \subfigure{\includegraphics[width=0.46\textwidth]{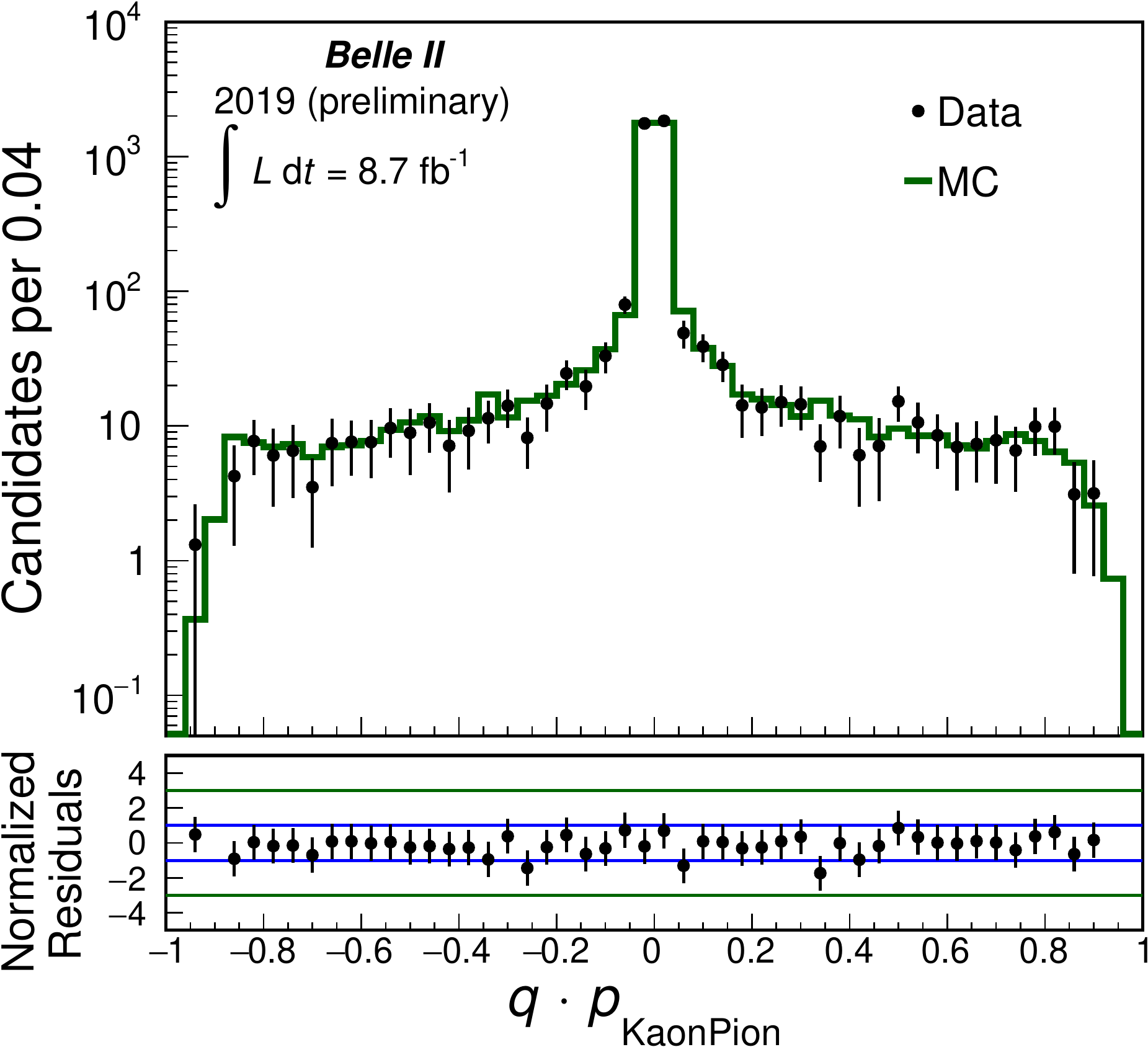}} 
    \subfigure{\includegraphics[width=0.46\textwidth]{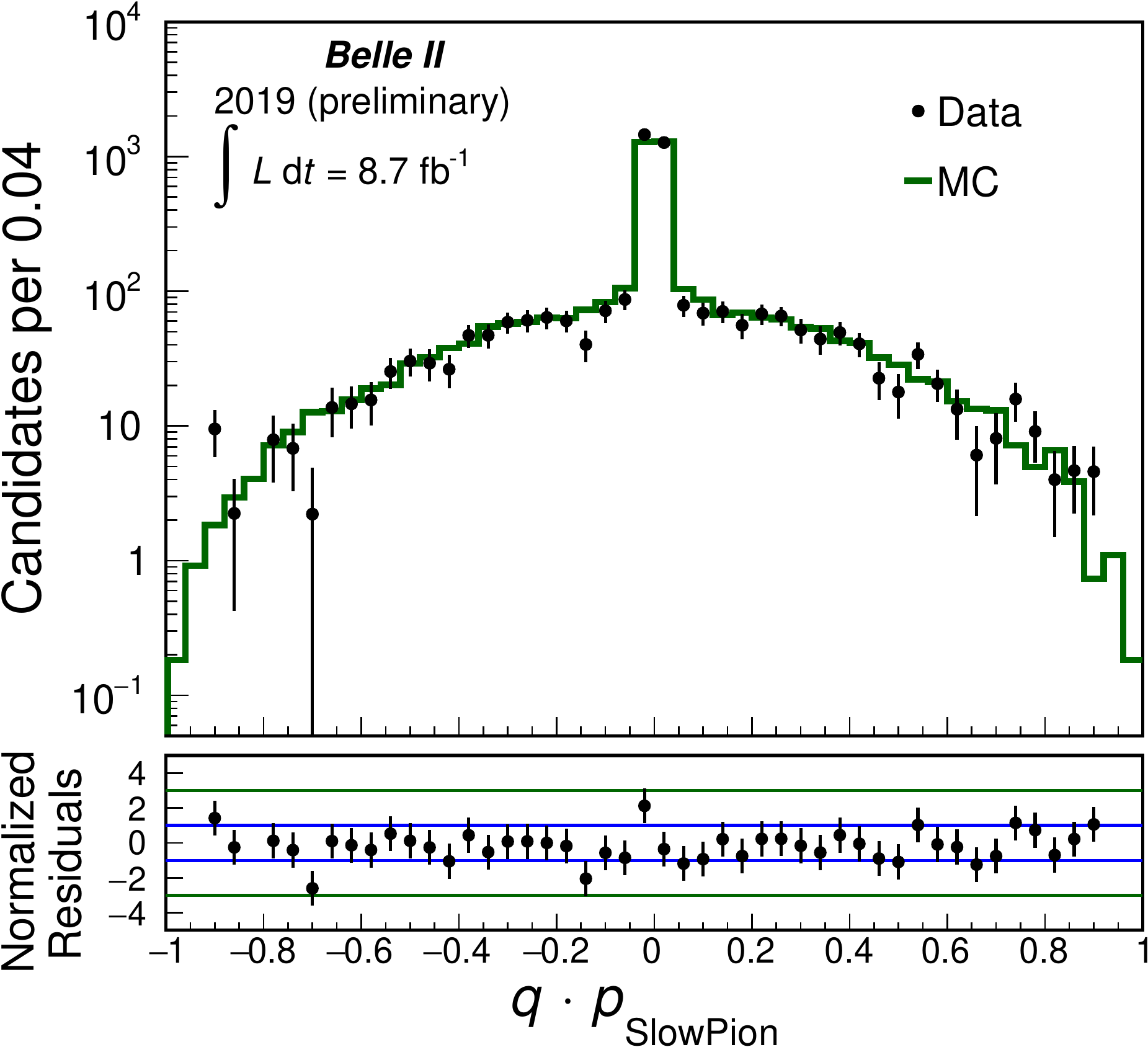}} \hfill 
    \subfigure{\includegraphics[width=0.46\textwidth]{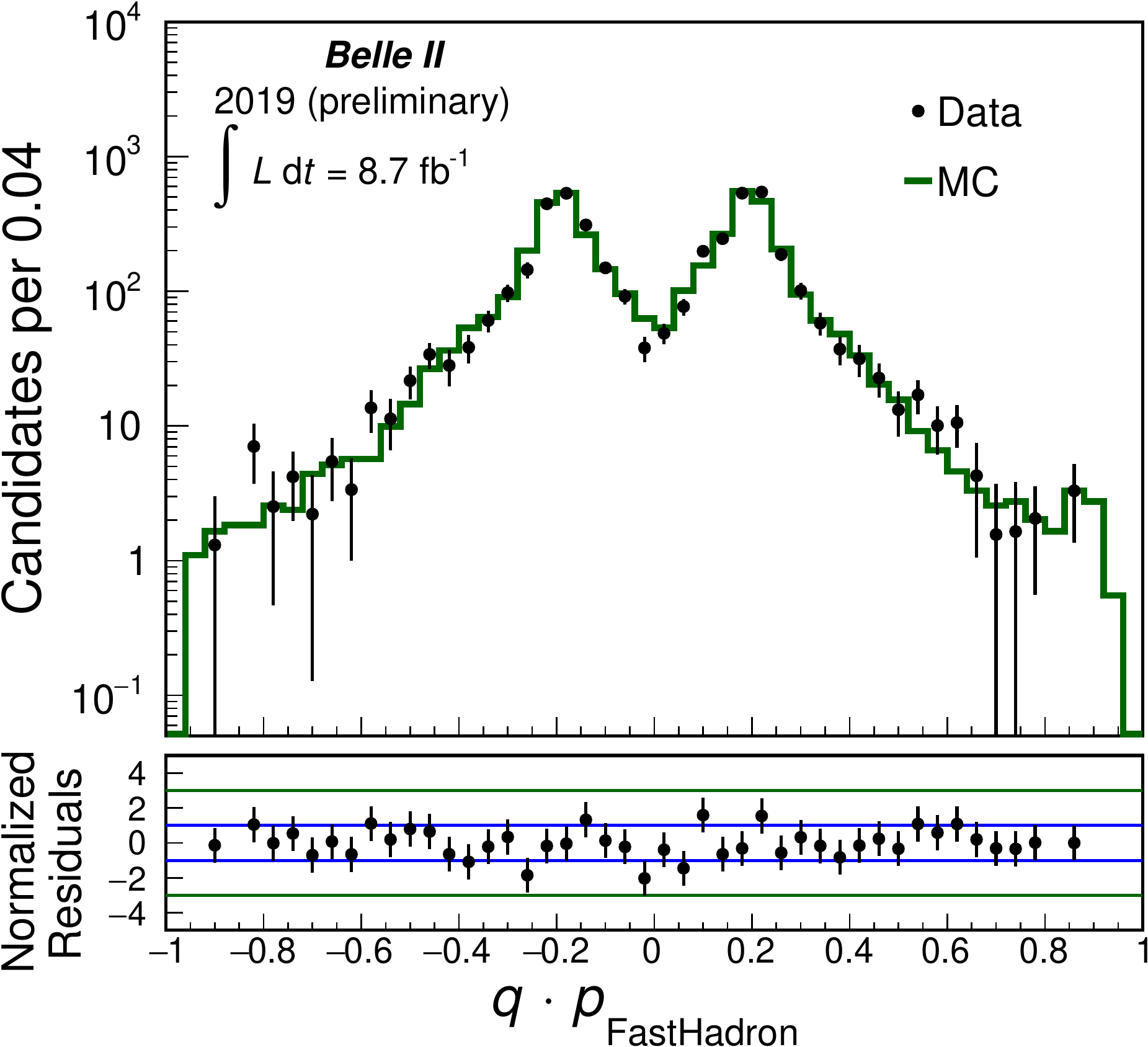}} 
    \subfigure{\includegraphics[width=0.46\textwidth]{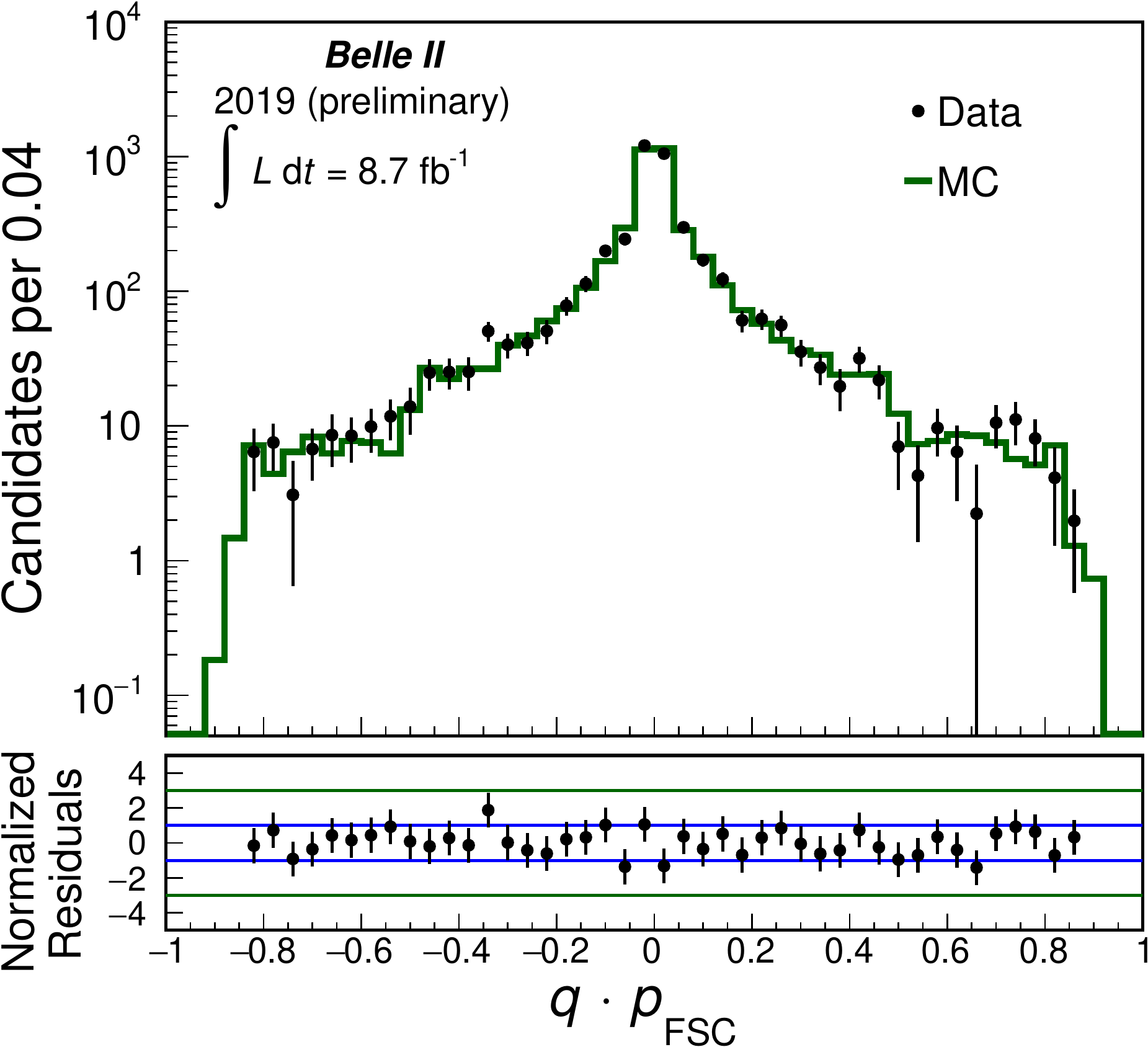}} \hfill
    \subfigure{\includegraphics[width=0.46\textwidth]{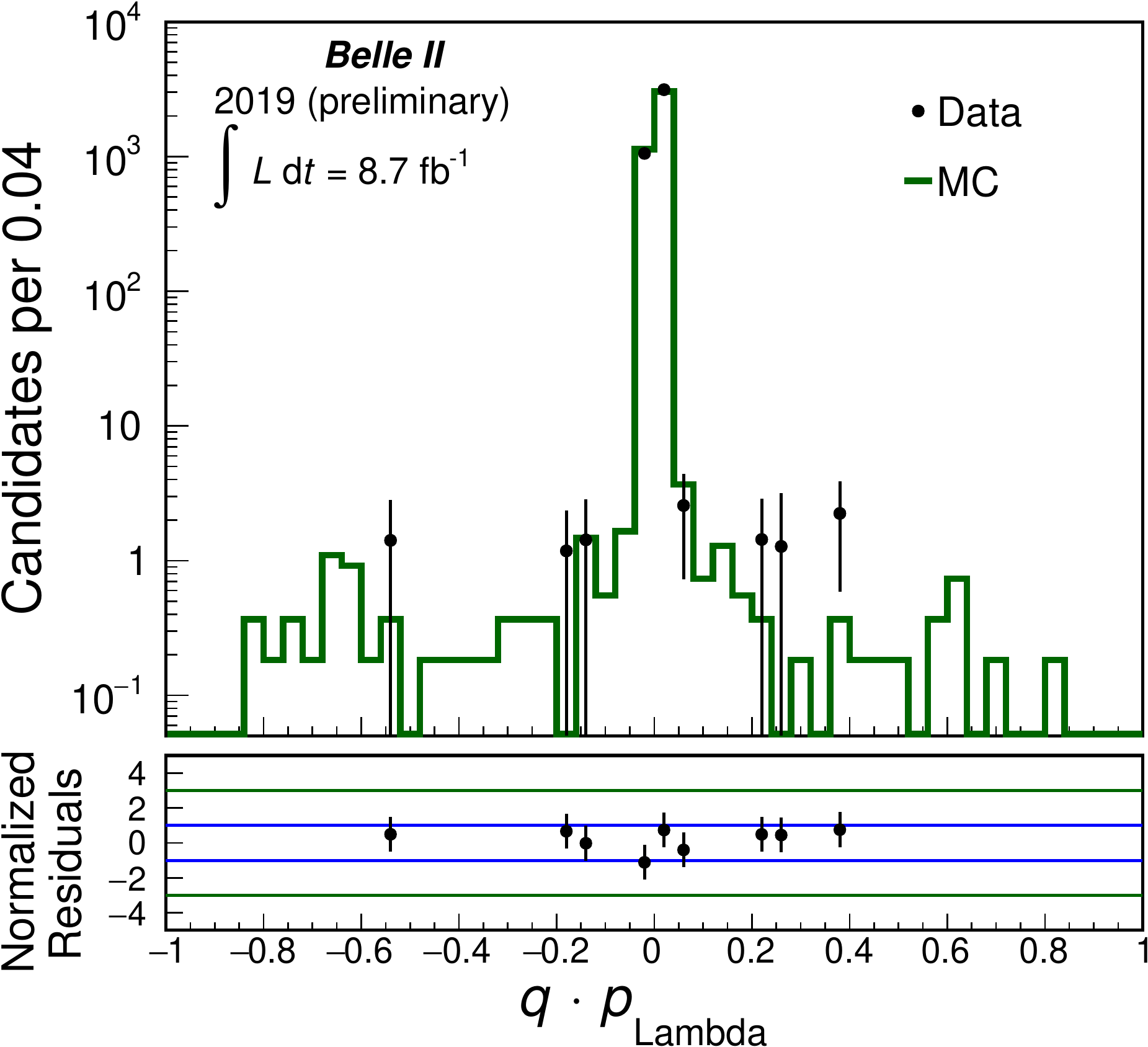}} 
    \caption{
    Normalized output distributions of individual tagging categories  in data and MC~\mbox{simulation} for $\PBzero\to\PD^{(*)-}\Ph^{+}$ candidates. The contribution from the signal component in data is compared with correctly associated signal MC~events~(2/3).}
    \label{fig:QP_data_splotMC_2}
\end{figure}

\begin{figure}
    \centering
    \subfigure{\includegraphics[width=0.46\textwidth]{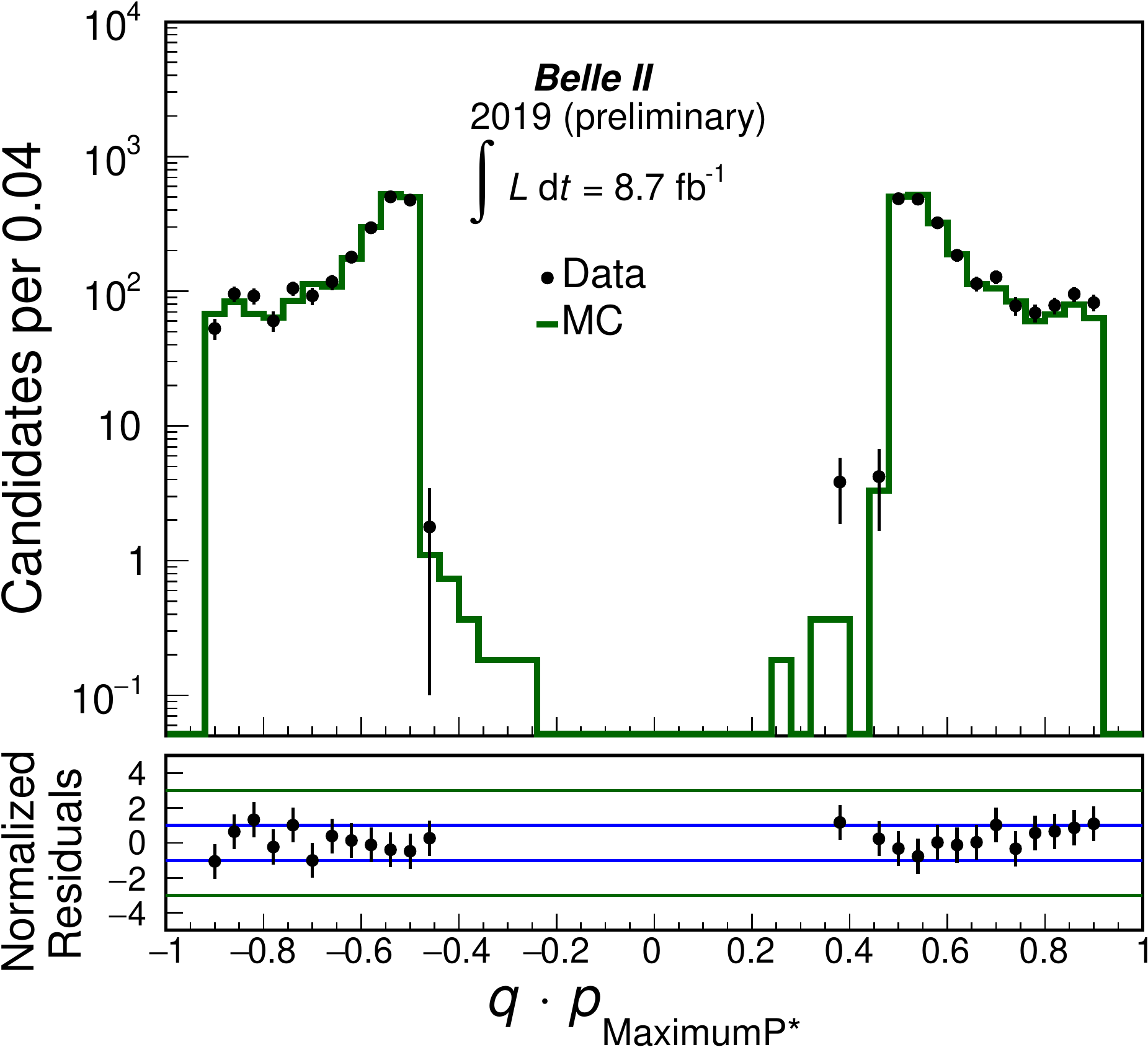}}
    \caption{ Normalized output distributions of individual tagging categories in data and MC~\mbox{simulation} for $\PBzero\to\PD^{(*)-}\Ph^{+}$ candidates. The contribution from the signal component in data is compared with correctly associated signal MC~events~(3/3).}
    \label{fig:QP_data_splotMC_3}
\end{figure}

Figures~\ref{fig:qrSigDistData} and~\ref{fig:qrBkgDistData} show the signal and background $q\cdot r$~distributions provided by the category-based flavor tagger for neutral and charged $\PB\to\PD^{(*)}\Ph^{+}$ candidates. We use the subindex ${\rm FBDT}$ to label the dilution provided by the flavor tagging algorithm.  We compare the signal data distribution with the distribution of correctly associated MC~events, and\linebreak the background data distributions with the distribution of side-band\linebreak MC~events~(\mbox{$M_{\rm bc} < \SI{5.27}{GeV}/c^2$} and same fit range \mbox{$\vert \Delta E \vert < \SI{0.12}{GeV}$}). We compare also the signal distributions in data and simulation for the individual tagging categories~\mbox{(Figures~\ref{fig:QP_data_splotMC_1}--\ref{fig:QP_data_splotMC_3})}. In general, the results show a good consistency between data and simulation.

\clearpage

\section{Results}

\label{sec:results}

Table~\ref{table:resFullDataFit} presents the results for the partial tagging efficiencies and the wrong-tag fractions obtained from the maximum-likelihood fit~(Sec.~\ref{sec:fullFit}) to data. To evaluate the tagging performance, we calculate the total effective efficiency as
\begin{equation*}
    \varepsilon_{\rm eff} =  \sum_{i} \varepsilon_{{\rm eff}, i} = \sum_{i} \varepsilon_i \cdot (1-2w_i)^2\text{,}
\end{equation*}
where $\varepsilon_{{\rm eff}, i}$ is the partial effective efficiency in the $i$-th $r$~bin. The effective tagging efficiency is a measure for the effective reduction of events due to the flavor dilution $r$. In \CP~violation analyses, the statistical uncertainty of measured \CP~asymmetries is approximately proportional to \mbox{$1/\sqrt{N_{\rm eff}}=1/\sqrt{N\cdot \varepsilon_{\rm eff}}$}, where $N_{\rm eff}$ is the number of effectively tagged events. Thus, one would obtain the same statistical precision for $N_{\rm eff}$ perfectly tagged events or for $N$ events tagged with an effective efficiency $\varepsilon_{\rm eff}$.

{\renewcommand{\arraystretch}{1.1}
\begin{table}[!h]
\centering
\caption{\label{table:resFullDataFit} Results of the maximum-likelihood fit to data: partial tagging efficiencies, wrong-tag fractions, partial effective efficiencies and total effective efficiency for neutral and charged \PB~candidates. The results are given with statistical and systematic uncertainties in percent.}
\vspace{5mm}
\begin{tabular}{ l  r  r  r  r}
\multicolumn{4}{l}{ $\PBzero\to\PD^{(*)-}h^{+}$}\\\hline
\hspace{0.3cm} $r$- Interval \hspace{0.5cm} &
\multicolumn{1}{r}{$\varepsilon_i\qquad\qquad\;\;$} &
\multicolumn{1}{r}{$w_i\qquad\qquad\;\;$}  &
\multicolumn{1}{c}{$\varepsilon_{\text{eff}, i}$}\\ \hline\hline
$ 0.000 - 0.100\quad$ & $ 20.3\pm 1.8 \pm 0.3\qquad$ & $ 47.4\pm 4.1 \pm 0.9\qquad $ & $   0.1 \pm 0.2 \pm 0.1\quad\, $ \\
$ 0.100 - 0.250\quad$ & $ 17.4\pm 0.9 \pm 0.1\qquad$ & $ 42.8\pm 4.4 \pm 0.5\qquad $ & $   0.4 \pm 0.4 \pm 0.1\quad\, $ \\
$ 0.250 - 0.500\quad$ & $ 21.2\pm 0.9 \pm 0.4\qquad$ & $ 26.9\pm 3.7 \pm 0.1\qquad $ & $   4.5 \pm 1.5 \pm 0.1\quad\, $ \\
$ 0.500 - 0.625\quad$ & $ 11.1\pm 0.7 \pm 0.2\qquad$ & $ 16.7\pm 5.0 \pm 2.4\qquad $ & $   4.9 \pm 1.5 \pm 0.7\quad\, $ \\
$ 0.625 - 0.750\quad$ & $ 9.6\pm 0.7 \pm 0.5\qquad$ & $ 9.2\pm 5.1 \pm 4.0\qquad $ & $   6.4 \pm 1.7 \pm 1.3\quad\, $ \\
$ 0.750 - 0.875\quad$ & $  7.0\pm 0.6 \pm 0.2\qquad$ & $ 12.0\pm 5.6 \pm 0.8\qquad $ & $   4.1 \pm 1.2 \pm 0.2\quad\, $ \\
$ 0.875 - 1.000\quad$ & $ 13.4\pm 0.7 \pm 0.3\qquad$ & $  0.0\pm 3.3 \pm 0.1\qquad $ & $  13.4 \pm 1.9 \pm 0.3\quad\, $ \\
\hline\hline
\multicolumn{1}{r}{Total} &  \multicolumn{3}{r}{ $\varepsilon_\text{eff} = \sum_i \varepsilon_i \cdot (1-2w_i)^2 =   33.8 \pm 3.6 \pm 1.6\quad\,$ } \\
\hline \multicolumn{4}{l}{ }\\
\multicolumn{4}{l}{ $\PBplus\to\APD^{(*)0}h^{+}$}\\
\hline
\hspace{0.3cm} $r$- Interval \hspace{0.5cm} &
\multicolumn{1}{r}{$\varepsilon_i\qquad\qquad\;\;$} &
\multicolumn{1}{r}{$w_i\qquad\qquad\;\;$}  &
\multicolumn{1}{c}{$\varepsilon_{\text{eff}, i}$}\\ \hline\hline
$ 0.000 - 0.100$ & $ 17.7\pm 1.7 \pm 0.4\qquad$ & $ 46.5\pm 2.7 \pm 0.4\qquad $ & $  0.1 \pm 0.1 \pm 0.1\quad\, $ \\ 
$ 0.100 - 0.250$ & $ 16.0\pm 0.8 \pm 0.2\qquad$ & $ 41.6\pm 2.7 \pm 1.6\qquad $ & $  0.5 \pm 0.3 \pm 0.2\quad\, $ \\ 
$ 0.250 - 0.500$ & $ 21.3\pm 0.9 \pm 0.1\qquad$ & $ 29.6\pm 2.1 \pm 0.9\qquad $ & $  3.6 \pm 0.8 \pm 0.3\quad\, $ \\ 
$ 0.500 - 0.625$ & $ 10.8\pm 0.7 \pm 0.2\qquad$ & $ 13.5\pm 2.6 \pm 0.8\qquad $ & $  5.8 \pm 0.9 \pm 0.3\quad\, $ \\ 
$ 0.625 - 0.750$ & $ 10.6\pm 0.7 \pm 0.5\qquad$ & $ 11.0\pm 2.3 \pm 0.7\qquad $ & $  6.5 \pm 0.9 \pm 0.4\quad\, $ \\ 
$ 0.750 - 0.875$ & $  9.1\pm 0.6 \pm 0.1\qquad$ & $  5.6\pm 1.8 \pm 0.2\qquad $ & $  7.2 \pm 0.7 \pm 0.1\quad\, $ \\ 
$ 0.875 - 1.000$ & $ 14.5\pm 0.6 \pm 0.4\qquad$ & $  2.8\pm 0.8 \pm 0.3\qquad $ & $ 12.9 \pm 0.7 \pm 0.4\quad\, $ \\
\hline\hline
\multicolumn{1}{r}{Total} &  \multicolumn{3}{r}{ $\varepsilon_\text{eff} = \sum_i \varepsilon_i \cdot (1-2w_i)^2 =   36.6 \pm 1.8 \pm 0.7\quad\,$ } \\
\hline
\end{tabular} 
\end{table}}

We consider systematic uncertainties associated with the model description, the $\Delta E$~fit range, the flavor mixing of the background,  the fit bias, and the bias introduced by peaking backgrounds.

\textbf{Model description:} we perform pseudo-experiments using an alternative model with a different $\Delta E$ parametrization. We perform 
fits to pseudo-data samples bootstrapped~(sampled with replacement) from the generic MC~simulation. We fit 
using the alternative and using the default model and calculate for each fit parameter~$x_i$ the difference~$\delta x_i$ between the results obtained with the alternative model and the results obtained with the default model. We obtain the mean difference $\delta \hat{x}_i$ by fitting a Gaussian function to the distribution of $\delta x_i$ and take the full mean $\delta \hat{x}_i$ as systematic uncertainty.
%For the alternative signal $\Delta E$ PDF, we use a double Gaussian function determined using truth-matched MC~events, with the additional flexibility of a global shift of peak position and a global scaling factor for the width. For the alternative background $\Delta E$ PDF, we use a first order polynomial function. The alternative model has the same amount of free fit parameters as the default one.

\textbf{$\Delta E$~Fit range:} Figures~\ref{fig:fit_dE_neutral_unbinned} and~\ref{fig:fit_dE_charged_unbinned} show that near the upper limit of the $\Delta E$~fit range there is an increase of the background that is slightly above the total fit model. We take into account possible systematic uncertainties due to this slight mismodeling near the upper limit by performing a fit in a reduced range \mbox{$-0.12 < \Delta E < 0.10\,\si{GeV}$}. We take as systematic uncertainty for each fit parameter $x_i$ the difference between the results obtained in the reduced range and the results obtained in the default $\Delta E$~range.

\textbf{Background mixing:} our fit takes into account the uncertainty on the world average for the signal $\chi_d$ in the Gaussian constraint. However, 
we assume that there is no mixing in the background~($\chi_d^{\rm bkg}=0$). Since the background includes $\PBzero\APBzero$ events, we study the effect of flavor mixing in the background by varying the value of the background $\chi_d^{\rm bkg}$ by a small amount $\pm \delta\chi_d^{\rm bkg}$, corresponding to the statistical uncertainty when we leave $\chi_d^{\rm bkg}$ free to float. We then take for each fit parameter~$x_i$ half the difference between the results for $\chi_d^{\rm bkg} + \delta\chi_d^{\rm bkg}$ and for  $\chi_d^{\rm bkg} - \delta\chi_d^{\rm bkg}$ as systematic uncertainty. %We obtain the value of $\delta\chi_d^{\rm bkg}$ by performing bootstrapped pseudo-experiments in which we leave $\chi_d^{\rm bkg}$ free to float and fix the signal $\chi_d$ to the true MC~value and the signal $\varepsilon_i$ and $w_i$ to the generated values~(count method). We take the width $\sigma$ of the residuals for $\chi_d^{\rm bkg}$ as $\delta\chi_d^{\rm bkg}$. 

\textbf{Fit\; bias:}  for each fit parameter $x_i$, we determine the fit bias using the residuals from bootstrapped pseudo-experiments. The residuals are the differences between the fit results for the individual pseudo-data samples and the fit results for the parent MC~sample. We take the full bias as systematic uncertainty.
   
\textbf{Peaking\; background bias:} we consider the bias caused by the peaking background, which is not included in the fit model, by calculating the difference between the results of the fit to the full MC~sample and the true values determined using MC~information. We take the full difference as systematic uncertainty.
   
We find the systematic uncertainty associated with the peaking background bias to be the dominant one  around $40\%$ of the statistical uncertainty, followed by the model description around $6\%$ and the fit bias around $3\%$.  The systematic uncertainties due to the fit range and due to the background mixing are around or below $1\%$ of the statistical uncertainty and therefore negligible. In future calibrations using larger data samples, we will consider the peaking background in the fit model and thus we expect the associated systematic uncertainty to decrease. With larger samples, we also expect to improve the fit model description of the data and thus to reduce the uncertainty due to the model description. 

\section{Linearity check}
\label{sec:linearity}

\begin{figure}
\centering
\includegraphics[width=0.6\linewidth]{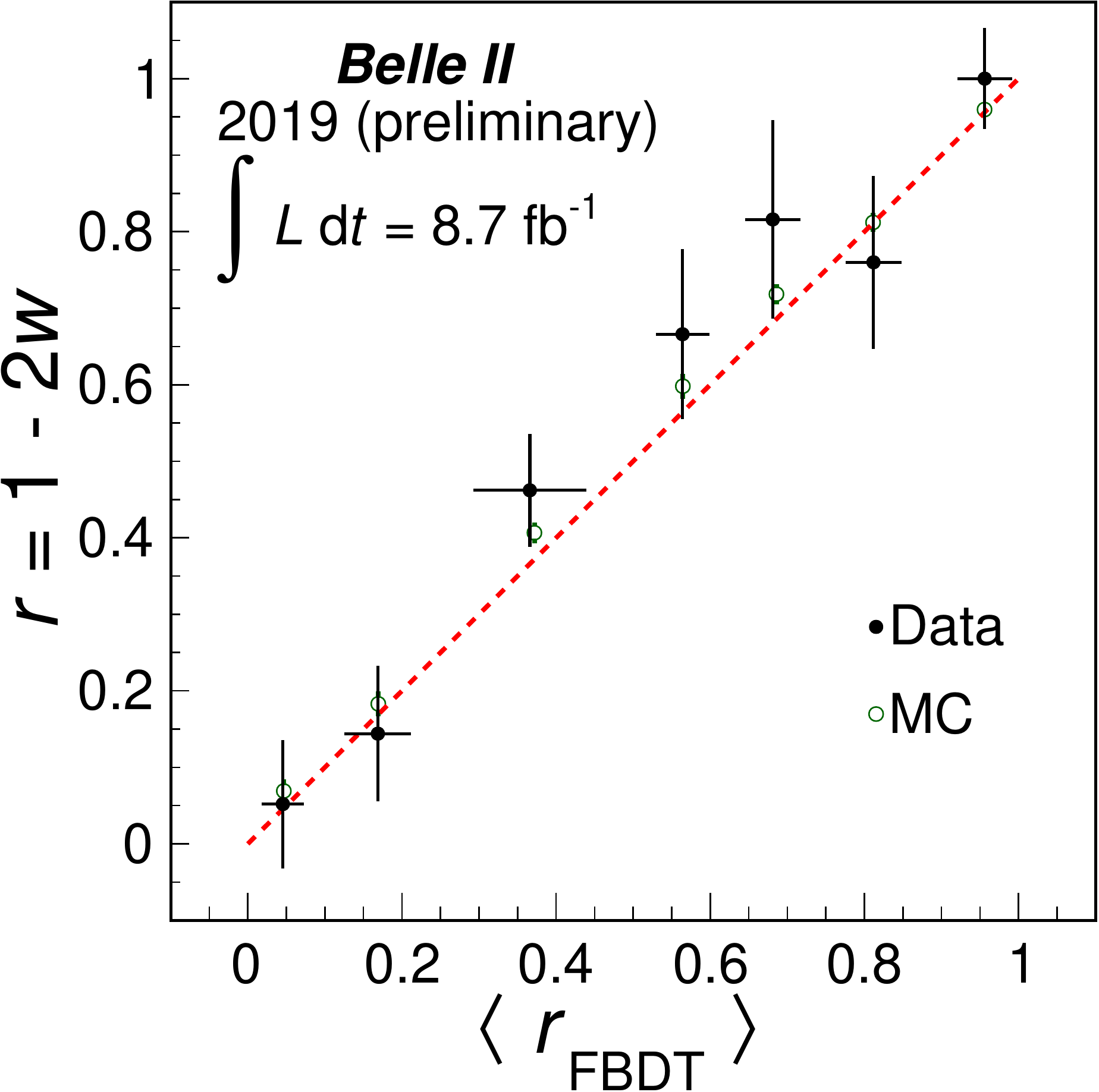}\\
\vspace{1cm}
\includegraphics[width=0.6\linewidth]{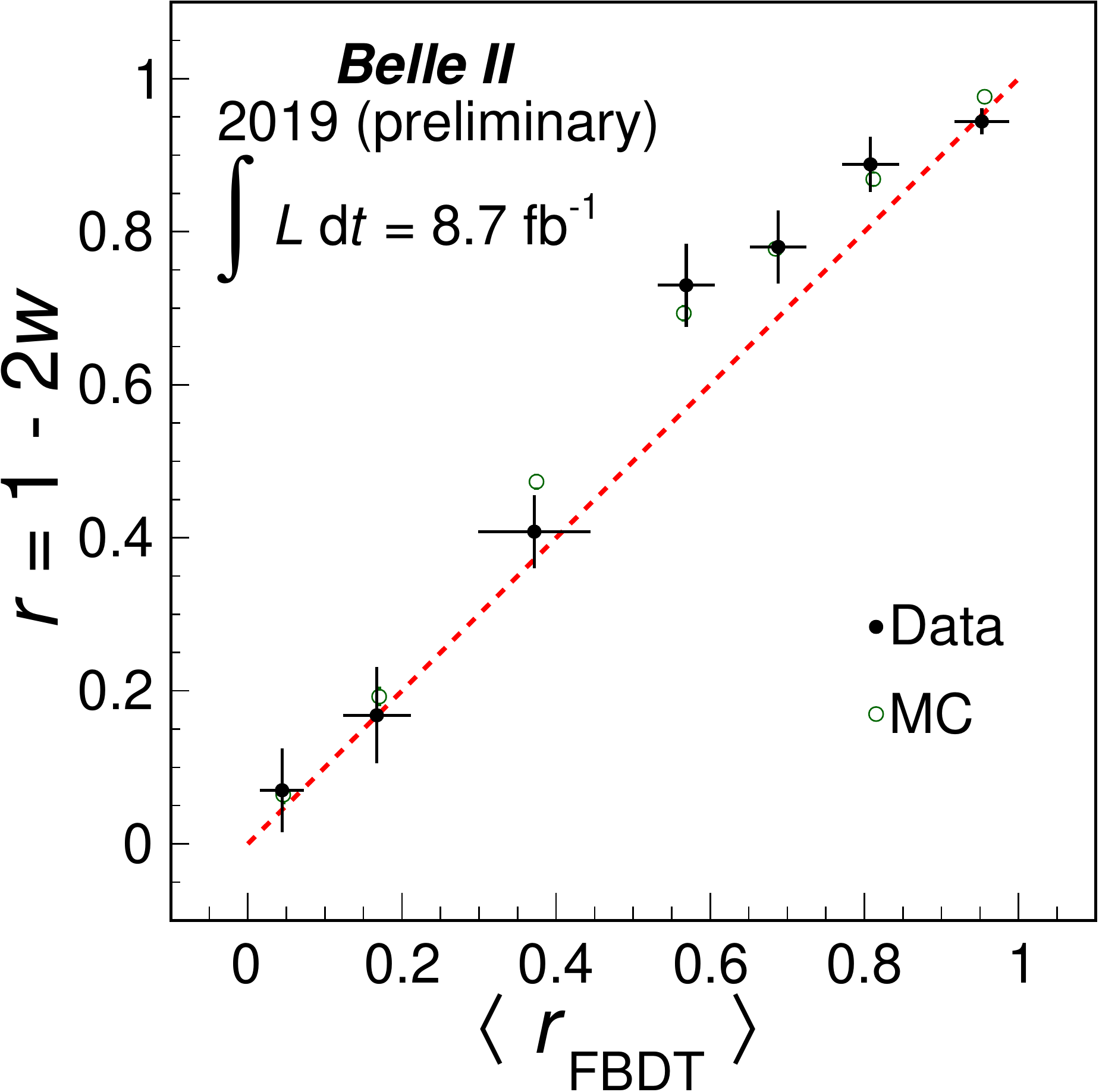}
\caption{\label{fig:cal_plot_data} Dilution factor $r = 1-2w$ as a function of the mean dilution~\mbox{$\langle \vert q\cdot r_{\rm FBDT} \vert \rangle$} provided by the flavor tagger in data and MC~simulation for (top)~neutral and (bottom)~charged \mbox{$\PB\to\PD^{(*)}\Ph^{+}$}~candidates. The red guidelines correspond to a linear function with an intercept at~$0$ and a slope of~$1$, i.e. to a perfect agreement between predicted and measured dilution. }
\end{figure}

By definition, the dilution factor $r$ is equal to $1-2w$. We probe if the dilution $r$ provided by the flavor tagger corresponds to the actual definition by performing a linearity check.  Figure~\ref{fig:cal_plot_data} shows the linearity check for simulation and data. 
For simulation, we determine the true wrong-tag fraction $w_{\rm MC}$ by comparing the MC~truth with the flavor tagger output, and calculate the true dilution $r_{\rm MC} = 1 -2w_{\rm MC}$. The mean dilution $\langle r_{\rm FBDT}\rangle$ is simply the mean of $\vert q\cdot r_{\rm FBDT}\vert$ for correctly associated MC~events in each $r$~bin.  For data, we obtain the mean \mbox{$\langle r_{\rm FBDT} \rangle = \langle\vert q\cdot r_{\rm FBDT} \vert\rangle$} values from the signal $q\cdot r_{\rm FBDT}$ distribution provided by the $s\mathcal{P}lot$ analysis in Sec.~\ref{sec:splot}. The dilution \mbox{$r=1-2\cdot w$} in data is obtained from the fit results for $w$. The linearity verifies the equivalence in average between the dilution provided by the flavor tagger and the measured one within the uncertainties.  For charged $\PB$~candidates, we observe a slightly non-linear behaviour which we attribute to the fact that the flavor tagger is optimized and trained only for neutral \PB~mesons. However, we observe a good agreement between data and simulation for both neutral and charged $\PB$~candidates.

\clearpage

\section{Comparison with Belle}

\label{sec:compBelle}

\begin{figure}
    \centering
    \includegraphics[width=0.9\textwidth]{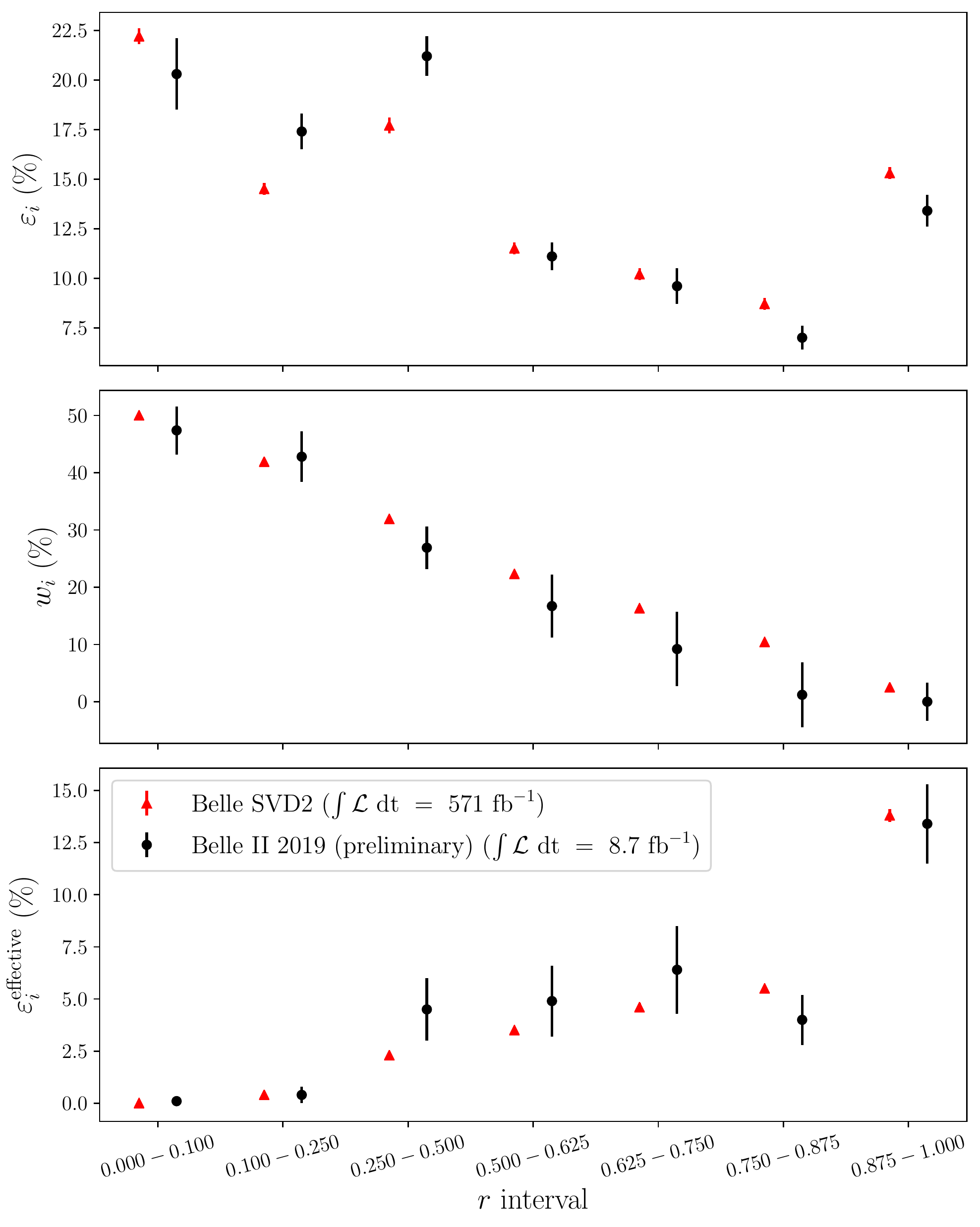}
    \caption{ Performance of the Belle~II flavor tagger in 2019 Belle~II data and of the Belle~flavor tagger in Belle data~\cite{Bevan:2014iga}  taken with the second silicon-vertex detector configuration~(SVD2).}
    \label{fig:Belle2VsBelle}
\end{figure}

Comparison of the current results with Belle's latest results~\cite{Bevan:2014iga} on flavor tagging provides interesting insight to assess Belle II's current and projected performance. We compare the wrong-tag fractions and the efficiencies in each $r$-bin, and the total effective efficiencies, which are shown in Table~\ref{tab:Belle2-vs-Belle} and Fig.~\ref{fig:Belle2VsBelle}.
The Belle flavor tagger reached an effective efficiency of~$(30.1\pm 0.4)\%$ on Belle data~\cite{Bevan:2014iga}.
In comparison with the previous Belle algorithm, the new Belle~II category-based flavor tagger considers more flavor signatures and more input variables, and is based on multivariate methods avoiding cut-based approaches.

{\renewcommand{\arraystretch}{1.1}
\begin{table}[!h]
\centering
\caption{\label{tab:Belle2-vs-Belle}Partial efficiencies $\varepsilon_i$ and wrong-tag fractions $w_i$ obtained with the Belle~II flavor tagger in 2019 Belle~II data and with the Belle flavor tagger in Belle data~\cite{Bevan:2014iga}  taken with the second silicon-vertex detector configuration~(SVD2). Statistical and systematical uncertainties are added in quadrature. All values are given in percent.}
\vspace{5mm}
\begin{tabular}{ l  r  r  r  r  r  r}
\multicolumn{1}{l}{ $\PBzero\to\PD^{(*)-}h^{+}$} & 
\multicolumn{2}{c}{$\varepsilon_i \pm \delta\varepsilon_{i}$} &
\multicolumn{2}{c}{$w_i \pm \delta w_i\;\, $} & \multicolumn{2}{c}{$\varepsilon_{\text{eff}, i} \pm \delta\varepsilon_{\text{eff}, i}\enskip\,$}\\\hline
$r$- Interval & \multicolumn{1}{c}{Belle~II} & \multicolumn{1}{c}{Belle}  & 
\multicolumn{1}{c}{Belle~II} & \multicolumn{1}{c}{Belle} & 
\multicolumn{1}{c}{Belle~II} & \multicolumn{1}{c}{Belle}\\ \hline\hline
$ 0.000 - 0.100$ & $20.3\pm 1.8$ & $22.2\pm 0.4$ & $47.4\pm 4.2$  & $50.0\quad $  &  $  0.1\pm 0.2$ &  $   0.0\quad$ \\
$ 0.100 - 0.250$ & $17.4\pm 0.9$ & $14.5\pm 0.3$ & $42.8\pm 4.4$  & $41.9\pm 0.4$ &  $  0.4\pm 0.4$ &  $  0.4\pm 0.1$ \\
$ 0.250 - 0.500$ & $21.2\pm 1.0$ & $17.7\pm 0.4$ & $26.9\pm 3.7$  & $31.9\pm 0.3$ &  $  4.5\pm 1.5$ &  $  2.3\pm 0.1$ \\
$ 0.500 - 0.625$ & $11.1\pm 0.7$ & $11.5\pm 0.3$ & $16.7\pm 5.5$  & $22.3\pm 0.4$ &  $  4.9\pm 1.7$ &  $  3.5\pm 0.1$ \\
$ 0.625 - 0.750$ & $ 9.6\pm 0.9$ & $10.2\pm 0.3$ & $ 9.2\pm 6.5$  & $16.3\pm 0.4$ &  $  6.4\pm 2.1$ &  $  4.6\pm 0.2$ \\
$ 0.750 - 0.875$ & $ 7.0\pm 0.6$ & $8.7 \pm 0.3$ & $ 1.2\pm 5.7$  & $10.4\pm 0.4$ &  $  4.0\pm 1.2$ &  $  5.5\pm 0.1$ \\
$ 0.875 - 1.000$ & $13.4\pm 0.8$ & $15.3\pm 0.3$ & $ 0.0\pm 3.3$  & $ 2.5\pm 0.3$ &  $ 13.4\pm 1.9$ &  $ 13.8\pm 0.3$ \\
\hline\hline
\multicolumn{2}{r}{Total} &  \multicolumn{3}{r}{ $\varepsilon_\text{eff} = \sum_i \varepsilon_i \cdot ( 1-2w_i)^2 =$} &  $33.8 \pm    3.9$  & $30.1\pm 0.4$\\
\hline
\end{tabular}
\end{table}
}

\clearpage

\section{Summary}
\label{sec:summary}

We report on the first calibration of the standard Belle~II~$\PB$-flavor tagger using 2019 Belle~II data. The $\Delta E$ distributions of reconstructed charmed $\PB$~candidates, restricted in $M_{\rm bc}$, are fit to identify the \PB~signals and measure the tagging efficiencies and the fractions of wrongly tagged events from the flavor evolution of the signal $\PB\APB$~pairs in a time-integrated way. The total effective efficiency for neutral $\PB$~candidates is measured to be 
\begin{center}
  \mbox{$\varepsilon_{\rm eff} = \big(33.8 \pm 3.6(\text{stat}) \pm 1.6(\text{sys})\big)\%$},   
\end{center}
and for charged $\PB$~candidates 
\begin{center}
   \mbox{$\varepsilon_{\rm eff} = \big(36.6 \pm 1.8(\text{stat}) \pm 0.7(\text{sys})\big)\%$}.  
\end{center}
The performance of the flavor tagger is generally compatible with expectations from \mbox{simulation}~(Fig.~\ref{fig:cal_plot_data}), establishing a good understanding of the detector performance. The performance is also comparable with the best one obtained by the Belle experiment within the uncertainties~(Fig.~\ref{fig:Belle2VsBelle}). This work marks a first milestone for future calibrations which will play an essential role in measurements of \CP-asymmetries.
%which will ultimately reveal or impose tight constraints on possible non-standard Model contributions. 

% \clearpage
% \begin{figure}[htb]
%  \centering
%  %\includegraphics[width=0.7\textwidth]{figures/hype_plot_fancy.pdf}
%  \includegraphics[width=0.7\textwidth]{figures/dE_BtoCharmless_hype_plot_fancy.pdf}
%  \caption{Stacked $\Delta E$ distributions of charmless channels reconstructed in 2019 Belle II data with the sum of fit projections overlaid.}
%  \label{fig:hype}
% \end{figure}

\section*{Acknowledgments}

We thank the SuperKEKB group for the excellent operation of the
accelerator; the KEK cryogenics group for the efficient
operation of the solenoid; and the KEK computer group for
on-site computing support.
This work was supported by the following funding sources:
%Armenia
Science Committee of the Republic of Armenia Grant No. 18T-1C180;
%Australia
Australian Research Council and research grant Nos.
DP180102629, 
DP170102389, 
DP170102204, 
DP150103061, 
FT130100303, 
and
FT130100018; 
%Austria
Austrian Federal Ministry of Education, Science and Research, and
Austrian Science Fund No. P 31361-N36; 
%Canada
Natural Sciences and Engineering Research Council of Canada, Compute Canada and CANARIE;
%China
Chinese Academy of Sciences and research grant No. QYZDJ-SSW-SLH011,
National Natural Science Foundation of China and research grant Nos.
11521505,
11575017,
11675166,
11761141009,
11705209,
and
11975076,
LiaoNing Revitalization Talents Program under contract No. XLYC1807135,
Shanghai Municipal Science and Technology Committee under contract No. 19ZR1403000,
Shanghai Pujiang Program under Grant No. 18PJ1401000,
and the CAS Center for Excellence in Particle Physics (CCEPP);
%Czech Republic
the Ministry of Education, Youth and Sports of the Czech Republic under Contract No.~LTT17020 and 
Charles University grants SVV 260448 and GAUK 404316;
%EU
European Research Council, 7th Framework PIEF-GA-2013-622527, 
Horizon 2020 Marie Sklodowska-Curie grant agreement No. 700525 `NIOBE,' 
and
Horizon 2020 Marie Sklodowska-Curie RISE project JENNIFER2 grant agreement No. 822070 (European grants);
%France
L'Institut National de Physique Nucl\'{e}aire et de Physique des Particules (IN2P3) du CNRS (France);
%Germany
BMBF, DFG, HGF, MPG, AvH Foundation, and Deutsche Forschungsgemeinschaft (DFG) under Germany's Excellence Strategy -- EXC2121 ``Quantum Universe''' -- 390833306 (Germany);
%India
Department of Atomic Energy and Department of Science and Technology (India);
%Israel
Israel Science Foundation grant No. 2476/17
and
United States-Israel Binational Science Foundation grant No. 2016113;
%Italy
Istituto Nazionale di Fisica Nucleare and the research grants BELLE2;
%Japan
Japan Society for the Promotion of Science,  Grant-in-Aid for Scientific Research grant Nos.
16H03968, 
16H03993, 
16H06492,
16K05323, 
17H01133, 
17H05405, 
18K03621, 
18H03710, 
18H05226,
19H00682, % Niigata
26220706,
and
26400255,
the National Institute of Informatics, and Science Information NETwork 5 (SINET5), 
and
the Ministry of Education, Culture, Sports, Science, and Technology (MEXT) of Japan;  
%Korea
National Research Foundation (NRF) of Korea Grant Nos.
2016R1\-D1A1B\-01010135,
2016R1\-D1A1B\-02012900,
2018R1\-A2B\-3003643,
2018R1\-A6A1A\-06024970,
2018R1\-D1A1B\-07047294,
2019K1\-A3A7A\-09033840,
and
2019R1\-I1A3A\-01058933,
Radiation Science Research Institute,
Foreign Large-size Research Facility Application Supporting project,
the Global Science Experimental Data Hub Center of the Korea Institute of Science and Technology Information
and
KREONET/GLORIAD;
%Malaysia
Universiti Malaya RU grant, Akademi Sains Malaysia and Ministry of Education Malaysia;
%Mexico
% CINVESTAV-IPN, UNAM, UAS, BUAP and CONACYT are funded under
Frontiers of Science Program contracts
FOINS-296,
CB-221329,
CB-236394,
CB-254409,
and
CB-180023, and SEP-CINVESTAV research grant 237 (Mexico);
%Poland
the Polish Ministry of Science and Higher Education and the National Science Center;
%Russia
the Ministry of Science and Higher Education of the Russian Federation,
Agreement 14.W03.31.0026;
%Saudi Arabia
University of Tabuk research grants
S-1440-0321, S-0256-1438, and S-0280-1439 (Saudi Arabia);
%Slovenia
Slovenian Research Agency and research grant Nos.
J1-9124
and
P1-0135; 
%Spain
Agencia Estatal de Investigacion, Spain grant Nos.
FPA2014-55613-P
and
FPA2017-84445-P,
and
CIDEGENT/2018/020 of Generalitat Valenciana;
%Taiwan
Ministry of Science and Technology and research grant Nos.
MOST106-2112-M-002-005-MY3
and
MOST107-2119-M-002-035-MY3, 
and the Ministry of Education (Taiwan);
%Thailand
Thailand Center of Excellence in Physics;
%Turkey
TUBITAK ULAKBIM (Turkey);
%Ukraine
Ministry of Education and Science of Ukraine;
%USA
the US National Science Foundation and research grant Nos.
PHY-1807007 % Luther
and
PHY-1913789, % Indiana CEEM
and the US Department of Energy and research grant Nos.
DE-AC06-76RLO1830, % PNNL
DE{}-SC0007983, % Wayne State
DE{}-SC0009824, % Florida
DE{}-SC0009973, % VPI
DE{}-SC0010073, % South Carolina
DE{}-SC0010118, % Carnegie Mellon
DE{}-SC0010504, % Hawaii
DE{}-SC0011784, % Cincinnati
DE{}-SC0012704; % BNL
%last group
and
%Vietnam
the National Foundation for Science and Technology Development (NAFOSTED) 
of Vietnam under contract No 103.99-2018.45.

\vspace{0.5cm}

\bibliography{belle2}
\bibliographystyle{belle2-note}
\end{document}